\newif\iffigs\figstrue
\documentclass[12pt,epsfig]{article}
\iffigs
 \usepackage{epsfig}
\else
\fi
\newcommand{\eqn}[1]{(\ref{#1})}

\newtheorem{propo}{Proposition}[section]
\newcommand{\bpropo}{\begin{propo}}
\newcommand{\epropo}{\end{propo}}
\newtheorem{teorema}{Theorem}[section]
\newcommand{\bth}{\begin{teorema}}
\newcommand{\eth}{\end{teorema}}
\newtheorem{prova}{Proof}[section]
\newcommand{\bpr}{\begin{prova}}
\newcommand{\epr}{\end{prova}}
\newsavebox{\uuunit}
\sbox{\uuunit}
    {\setlength{\unitlength}{0.825em}
     \begin{picture}(0.6,0.7)
        \thinlines
        \put(0,0){\line(1,0){0.5}}
        \put(0.15,0){\line(0,1){0.7}}
        \put(0.35,0){\line(0,1){0.8}}
       \multiput(0.3,0.8)(-0.04,-0.02){12}{\rule{0.5pt}{0.5pt}}
     \end {picture}}

\def\IP{\relax{\rm I\kern-.18em P}}

\begin{document}
%
\font\cmss=cmss10 \font\cmsss=cmss10 at 7pt
\def\twomat#1#2#3#4{\left(\matrix{#1 & #2 \cr #3 & #4}\right)}
\def\inbar{\vrule height1.5ex width.4pt depth0pt}
\def\IC{\relax\,\hbox{$\inbar\kern-.3em{\rm C}$}}
\def\IG{\relax\,\hbox{$\inbar\kern-.3em{\rm G}$}}
\def\IB{\relax{\rm I\kern-.18em B}}
\def\ID{\relax{\rm I\kern-.18em D}}
\def\IL{\relax{\rm I\kern-.18em L}}
\def\IF{\relax{\rm I\kern-.18em F}}
\def\IH{\relax{\rm I\kern-.18em H}}
\def\II{\relax{\rm I\kern-.17em I}}
\def\IN{\relax{\rm I\kern-.18em N}}
\def\IP{\relax{\rm I\kern-.18em P}}
\def\IQ{\relax\,\hbox{$\inbar\kern-.3em{\rm Q}$}}
\def\bfzero{\relax\,\hbox{$\inbar\kern-.3em{\rm 0}$}}
\def\IK{\relax{\rm I\kern-.18em K}}
\def\IG{\relax\,\hbox{$\inbar\kern-.3em{\rm G}$}}
 \font\cmss=cmss10 \font\cmsss=cmss10 at 7pt
\def\IR{\relax{\rm I\kern-.18em R}}
\def\ZZ{\relax\ifmmode\mathchoice
{\hbox{\cmss Z\kern-.4em Z}}{\hbox{\cmss Z\kern-.4em Z}}
{\lower.9pt\hbox{\cmsss Z\kern-.4em Z}}
{\lower1.2pt\hbox{\cmsss Z\kern-.4em Z}}\else{\cmss Z\kern-.4em
Z}\fi}
\def\bfone{\relax{\rm 1\kern-.35em 1}}
\def\dop{{\rm d}\hskip -1pt}
\def\real{{\rm Re}\hskip 1pt}
\def\trace{{\rm Tr}\hskip 1pt}
\def\ii{{\rm i}}
\def\diag{{\rm diag}}
\def\sch#1#2{\{#1;#2\}}
\def\bfone{\relax{\rm 1\kern-.35em 1}}
\font\cmss=cmss10 \font\cmsss=cmss10 at 7pt
\def\a{\alpha} \def\b{\beta} \def\d{\delta}
\def\e{\epsilon} \def\c{\gamma}
\def\G{\Gamma} \def\l{\lambda}
\def\L{\Lambda} \def\s{\sigma}
\def\cA{{\cal A}} \def\cB{{\cal B}}
\def\cC{{\cal C}} \def\cD{{\cal D}}
\def\cF{{\cal F}} \def\cG{{\cal G}}
\def\cH{{\cal H}} \def\cI{{\cal I}}
\def\cJ{{\cal J}} \def\cK{{\cal K}}
\def\cL{{\cal L}} \def\cM{{\cal M}}
\def\cN{{\cal N}} \def\cO{{\cal O}}
\def\cP{{\cal P}} \def\cQ{{\cal Q}}
\def\cR{{\cal R}} \def\cV{{\cal V}}\def\cW{{\cal W}}
\newcommand{\be}{\begin{equation}}
\newcommand{\ee}{\end{equation}}
\newcommand{\bea}{\begin{eqnarray}}
\newcommand{\eea}{\end{eqnarray}}
\let\la=\label \let\ci=\cite \let\re=\ref
%
%
%
\def\crr{\crcr\noalign{\vskip {8.3333pt}}}
\def\tilde{\widetilde}
\def\bar{\overline}
\def\us#1{\underline{#1}}
\let\shat=\hat
\def\hat{\widehat}
\def\hyp{\vrule height 2.3pt width 2.5pt depth -1.5pt}
\def\square{\mbox{.08}{.08}}
\def\Coeff#1#2{{#1\over #2}}
\def\Coe#1.#2.{{#1\over #2}}
\def\coeff#1#2{\relax{\textstyle {#1 \over #2}}\displaystyle}
\def\coe#1.#2.{\relax{\textstyle {#1 \over #2}}\displaystyle}
\def\half{{1 \over 2}}
\def\shalf{\relax{\textstyle {1 \over 2}}\displaystyle}
\def\dag#1{#1\!\!\!/\,\,\,}
\def\to{\rightarrow}
\def\notin{\hbox{{$\in$}\kern-.51em\hbox{/}}}
\def\shdot{\!\cdot\!}
\def\ket#1{\,\big|\,#1\,\big>\,}
\def\bra#1{\,\big<\,#1\,\big|\,}
\def\equaltop#1{\mathrel{\mathop=^{#1}}}
\def\Trbel#1{\mathop{{\rm Tr}}_{#1}}
\def\inserteq#1{\noalign{\vskip-.2truecm\hbox{#1\hfil}
\vskip-.2cm}}
\def\attac#1{\Bigl\vert
{\phantom{X}\atop{{\rm\scriptstyle #1}}\phantom{X}}}
\def\exx#1{e^{{\displaystyle #1}}}
\def\del{\partial}
\def\delbar{\bar\partial}
\def\nex#1{$N\!=\!#1$}
\def\dex#1{$d\!=\!#1$}
\def\cex#1{$c\!=\!#1$}
\def\eg{{\it e.g.}} \def\ie{{\it i.e.}}
\def\IE{\relax{{\rm I\kern-.18em E}}}
\def\cE{{\cal E}}
\def\rt{{\cR^{(3)}}}
\def\IGam{\relax{{\rm I}\kern-.18em \Gamma}}
\def\IGa{\IA}
\def\ii{{\rm i}}
\begin{titlepage}
\begin{center}
{\LARGE    N=8  gaugings revisited :  \\
\vskip 0.2cm
an exhaustive classification$^{*\dagger}$
}\\
\vfill
{\large Francesco Cordaro$^1$, Pietro Fr\'e$^1$, Leonardo Gualtieri$^1$,
\\
\vskip 0.2cm
Piet Termonia$^1$, and  Mario Trigiante$^2$   } \\
\vfill
{\small
$^1$ Dipartimento di Fisica Teorica, Universit\'a di Torino, via P. Giuria 1,
I-10125 Torino, \\
 Istituto Nazionale di Fisica Nucleare (INFN) - Sezione di Torino, Italy \\
\vspace{6pt}
$^2$ Department of Physics, University of Wales Swansea, Singleton
Park,\\
 Swansea SA2 8PP, United Kingdom\\
\vspace{6pt}
}
\end{center}
\vfill
\begin{center}
{\bf Abstract}
\end{center}
{\small
In this paper we reconsider, for $N=8$ supergravity,
the problem of gauging the most general electric subgroup.
 We show that admissible theories
are fully characterized by a single algebraic equation to be satisfied
by the embedding ${\cal G}_{gauge} \rightarrow  SL(8,\IR) \subset E_{7(7)}$.
The complete set of solutions to this equation   contains
$36$ parameters. Modding by the action of $SL(8,\IR)$ conjugations that yield equivalent theories
all continuous parameters are eliminated except for an overall coupling constant
and we obtain a discrete set of orbits. This set
is in one--to--one correspondence with $36$ Lie subalgebras of $SL(8,\IR)$, corresponding to all
possible   real forms of the $SO(8)$ Lie algebra plus a set of contractions thereof.
By means of our analysis we establish the theorem that the N=8 gaugings
constructed by Hull in the middle eighties constitute the exhaustive
set of models. As a corollary we show that there exists a unique $7$--dimensional abelian gauging.
The corresponding abelian algebra is not contained in the maximal abelian ideal
of the solvable Lie algebra generating the scalar manifold $E_{7(7)}/SU(8)$ }
\vspace{2mm} \vfill \hrule width 3.cm
{\footnotesize
 $^*$ Supported in part by   EEC  under TMR contract
 ERBFMRX-CT96-0045\\
 $^{\dagger}$ Supported in part by EEC  under TMR contract
 ERBFMRXCT960012, in which M. Trigiante is associated to Swansea
University}
\end{titlepage}
\section{Introduction}
Supergravity theory  was  developed in the late seventies
and in the early eighties in a completely independent way from string theory.
Yet it has proved to encode a surprising wealth of non perturbative information about
the various superstrings and the  candidate microscopic theory that unifies
all of them: M--theory. \cite{secrevol1,secrevol2} Most prominent about the features of
supergravity models that teach us something about non perturbative
string states is the U--duality group, namely the isometry group of
the scalar sector whose discrete subgroup $U(\ZZ)$ is believed to be
an exact symmetry of the non--perturbative string spectrum
\cite{secrevol1}.
This viewpoint has been successfully applied to the construction of
BPS saturated $p$--brane  solutions that provide the missing partners
of perturbative string states needed to complete U--duality multiplets
\cite{mbrastelle,mbratownsend}. In particular in space--time dimension $D=4$ an active
study has been performed of BPS $0$--branes, namely extremal
black--holes, for extended supergravities with $ 2 \le N \le 8$
\cite{ferkal,kallosh,doubling,noi3}.
These non--perturbative quantum states are described by classical
solutions of the ungauged version of supergravity which in the first
two years after the second string revolution \cite{secrevol1,secrevol2} has been the
focus of attention of string theorists.  {\it Ungauged supergravity}
corresponds to the case where the maximal symmetric solution is
Minkowski space $M^{Mink}_D$ and furthermore no field in the theory is charged with
respect to the vector fields, in particular the {\it graviphotons}.
\par
Another aspect of supergravity that has recently received renewed
attention is its {  gauging}.  {\it Gauged supergravity} corresponds
to the situation where the maximal symmetric solution is not
necessarily Minkowski space but can also be anti--de Sitter space $AdS_D$ and where
fields are charged with respect to the vectors  present in the
theory, in particular the graviphotons (for a review see \cite{mylecture}).
In gauged supergravity three modifications occur:
\begin{enumerate}
\item A subgroup $G_{gauge} \subset G_{elect} \subset U$ is gauged,
namely becomes a local symmetry. $G_{elect}$ is by definition the
subgroup of the U--duality group that maps the set of electric field
strengths into itself without mixing them with the magnetic ones. As
a consequence U--duality is classically broken and only its discrete
part $U(\ZZ)$ can survive non--perturbatively.
\item The supersymmetry transformation rules are modified through the
addition of non--derivative terms that depend only on the scalar
fields and that are usually named the {\it fermion  shifts}
$\Sigma_A^i(\phi)$. Namely we have:
 \begin{equation}
\delta  \mbox{fermion}^i = \dots \, + \, \Sigma_A^i(\phi) \, \epsilon^A
\end{equation}
\item A scalar potential ${\cal V}(\phi)$ is generated that is
related to the fermion shifts by a bilinear Ward identity \cite{Wardide}
\begin{equation}
 \delta_A^B \,  {\cal V}(\phi) \, = \, \Sigma_A^i(\phi) \, \Sigma^{Bj}(\phi) \, M_{ij}
\end{equation}
where $M_{ij}$ is a suitable constant matrix depending on the
specific supergravity model considered and where the lower and upper
capital latin indices $A,B,\dots$ enumerate the left and right
chiral projection of the supersymmetry charges, respectively.
\end{enumerate}
Recently it has been verified \cite{goverh,doubling,popelu} that one can construct $Mp$--brane   and
$Dp$--brane solutions of either D=11 \cite{sugra11a,sugra11b} or type IIB D=10 supergravity \cite{sugra10b}
with the following property: near the horizon their structure is
described by a {\it gauged supergravity theory} in dimension $d=p+2$ and
the near horizon geometry \cite{leuvenpap} is:
\begin{equation}
{\cal M}^h = AdS_{p+2} \, \times \, {M}_{D-p-2}
\label{adsxm}
\end{equation}
$AdS_{p+2}$ being a notation for anti--de Sitter space and ${M}_{D-p-2}$ denoting an Einstein
manifold in the complementary dimensions.  Hence near the horizon one
has a Kaluza--Klein expansion on ${M}_{D-p-2}$ whose truncation to zero--modes is
gauged supergravity, the extension $N$  being decided
by the number of Killing spinors admitted by ${M}_{D-p-2}$ and the
group $G_{gauge}$ that is gauged being related to the isometry group $G_{iso}$ of this manifold.
For instance the simplest example of $M2$ brane flows at the horizon to the geometry
\begin{equation}
{\cal M}^h = AdS_{4} \, \times \, {S}^{7}
\end{equation}
and in that region it is described by the $SO(8)$ gauging of $N=8$
supergravity, namely the 
de Wit and Nicolai theory \cite{dwni}.  Furthermore, as it is
extensively discussed in recent literature one finds a duality
between Kaluza--Klein supergravity yielding the gauged model and the
conformal field theory on the world volume describing the microscopic
degrees of freedom of the quantum brane.
\par
In view of this relation an important problem that emerges is that of
determining the precise relation between the various possible
gaugings of the same ungauged supergravity theory and the near
horizon geometry of various branes.  In this paper we do not attempt
to solve such a problem but we have it as a clear motivation  for our
revisitation of the problem we actually address: {\it the  classification of  all possible
gaugings for $N=8$ supergravity.}
\par
Another reason for revisiting the gauging problem comes from the
lesson taught us by the $N=2$ case.  In \cite{lucianoi},
extending work of \cite{fergir}, we discovered that
$N=2$ supersymmetry can be spontaneously broken to $N=1$ when the
following conditions are met:
\begin{itemize}
\item The scalar manifold of supergravity, which is generically given
by the direct product ${\cal SK} \otimes {\cal QK}$ of a special K\"ahler
manifold with a quaternionic one, is a {\it homogeneous non--compact
coset manifold} ${\cal G}/{\cal H}$
\item Some {\it translational abelian} symmetries of ${\cal G}/{\cal H}$
are gauged.
\end{itemize}
The basic ingredient in deriving the above result is the Alekseevskian
description \cite{alex} of the scalar manifold ${\cal SK} \otimes {\cal QK}$ in
terms of {\it solvable} Lie algebras, a K\"ahler algebra ${\cal K}$ for the vector
multiplet sector ${\cal SK}$ and a quaternionic algebra ${\cal Q}$
for the hypermultiplet sector ${\cal QK}$. By means of this
description the homogeneous non compact coset manifold ${\cal G}/{\cal H}$
is identified with the {\it solvable group manifold} $\exp[Solv]$
where
\begin{equation}
  Solv  = {\cal K} \, \oplus  \, {\cal Q}  \label{k+q}
\end{equation}
and the translational symmetries responsible for the supersymmetry
breaking are identified with suitable abelian subalgebras
\begin{equation}
{\cal T} \, \subset \, Solv     \label{tinsolv}
\end{equation}
An obvious observation that easily occurs once such a perspective is
adopted is the following one: for all extended supergravities with
$N \ge 3$ the scalar manifold is a   homogeneous non--compact
coset manifold  ${\cal G}/{\cal H}$. Hence it was very tempting to extend
the {\it solvable Lie algebra approach} to such supergravity theories, in
particular to the maximal extended ones in all dimensions $4 \le D
\le 10$. This is what was done in the series of three papers
\cite{noi1,noi2,noi3}. Relying on a well established mathematical theory
which is available in standard textbooks (for instance
\cite{helgason}), every non--compact homogeneous space ${\cal G}/{\cal H}$
is  a solvable group manifold and its generating solvable Lie
algebra $Solv \left( {\cal G}/{\cal H} \right)$ can be constructed
utilizing roots and Dynkin diagram techniques. This fact
offers the so far underestimated possibility of introducing an
intrinsic algebraic characterization of the supergravity scalars. In
relation with string theory this yields
a group--theoretical definition of Ramond  and Neveu--Schwarz scalars.
In the case of $N=8$ supergravity  the scalar coset manifold $E_{7(7)}/SU(8)$
is generated by   a solvable Lie algebra with the following structure:
\begin{equation}
Solv(E_{7(7)})={\cal H}_7 \oplus \Phi^{+}(E_{7})
\label{decompo1}
\end{equation}
where ${\cal H}_7$ is the Cartan subalgebra and
where $\Phi^{+}(E_{7})$ is the $63$ dimensional positive part of the $E_7$ root space.
As shown in \cite{noi2}  this latter admits the decomposition:
\begin{equation}
\Phi^{+}(E_7)=\Phi^+(E_2) \oplus \ID^{+}_{2} \oplus \ID^{+}_{3}
\oplus \ID^{+}_{4} \oplus \ID^{+}_{5}  \oplus \ID^{+}_{6}
\label{decompo}
\end{equation}
where $ \Phi^+(E_2) $ is the one--dimensional root space of the
U--duality group in $D=9$ and $\ID^{+}_{r+1} = {\cal A}_{r+2}$ 
the {\it maximal abelian ideal} of the
U--duality solvable Lie algebra in $D=10-r-1$ dimensions. In particular for the $N=8,D=4$ theory the
maximal abelian ideal of the solvable Lie algebra is:
\begin{equation}
\begin{array}{rclcl}
\mbox{max. abel. ideal} & \rightarrow & {\cal A}_{7} & \equiv & \ID^{+}_{6}   \subset  Solv(E_{7(7)})  \cr
\mbox{dim} {\cal A}_{7} & = & 27 & \null &  \null \cr
\end{array}
\end{equation}
Therefore, relying on the experience gained in the context of the $N=2$ case
in \cite{noi2} we have considered the possible gauging of an
abelian subalgebra ${\cal A}_{gauge}$ within this maximal abelian
ideal. In this context by {\it maximal} we refer to the property of having 
{\it maximal dimension}. It was reasonable to expect that such a gauging might
produce a spontaneous breaking of supersymmetry with flat directions
just as it happened in the lower supersymmetry example.
The basic consistency criterion to
gauge an $n$--dimensional subgroup of the translational
symmetries is that we find at least $n$ vectors which are  inert under the
action of the proposed subalgebra. Indeed the
  set of vectors that can gauge an abelian algebra
(being in its adjoint representation) must be neutral under the
action of such an algebra.  In \cite{noi2}, applying this
criterion we found that the   gaugeable subalgebra of the
maximal abelian ideal is $7$--dimensional.
\begin{equation}
7 = \mbox{dim} {\cal A}_{gaugeable} \, \subset \, {\cal A}_7 \,
\subset \, Solv(E_{7(7)})
\end{equation}
Yet this was only a necessary but not sufficient criterion for the
existence of the gauging. So whether translationally isometries of
the solvable Lie algebra could be gauged was not established in
\cite{noi2}.
\par
The present paper provides an answer to this question and the answer
is negative. Namely, we derive an algebraic equation that yields {\it a
necessary and sufficient} condition for $N=8$ gaugings and we obtain
the complete set of its solutions. This leads to an exhaustive
classifications of all gauged $N=8$ supergravities. The set we obtain
coincides with the set of gauged models obtained more than ten years
ago by Hull \cite{hull} whose completeness was not established, so far.
In the classification there appears a unique abelian gauging and in
this case the gauged Lie algebra is indeed $7$--dimensional. However
this abelian algebra, named $CSO(1,7)$ in Hull's terminology is not
contained in the maximal abelian ideal ${\cal A}_7$ since 
one of its elements belongs to $\ID_5^+$.\par
So a question that was left open in \cite{noi2} is answered by
the present paper.
\par
Having established the result we present in this paper the next step
that should be addressed is the study of critical points for all the
classified gaugings, their possible relation with $p$--brane
solutions and with partial supersymmetry breaking possibly in
$AdS_4$, rather than in Minkowski space. Such a program will be
undertaken elsewhere.
\par
The rest of the paper is devoted to proving the result we have
described.
\section{N=8 Gauged Supergravity}
In this section we summarize the structure of gauged N=8 supergravity
and we pose the classification problem we have solved.
\subsection{The bosonic lagrangian}
According to the normalizations of
\cite{n2paperone} and \cite{noi3} the bosonic Lagrangian of gauged
N=8 supergravity has the following form:
\begin{equation}
\begin{array}{ccl}
{\cal L} & = &\sqrt{-g} \, \Bigl ( 2\, R[g] + \frac{1}{4}\, \mbox{Im}\,
{\cal N}_{\Lambda\Sigma, \Delta\Pi} \,
{\cal F}^{\Lambda\Sigma\vert \mu \nu}\,{\cal
F}^{\Delta \Pi}_{\mu \nu}
+ \frac{1}{4}
\mbox{Re}\, {\cal N}_{\Lambda\Sigma,\Delta\Pi} \,
{\cal F}^{\Lambda\Sigma}_{\mu \nu} \, {\cal
F}^{\Delta\Pi}_{\rho \sigma}\, \epsilon^{\mu \nu\rho\sigma}
\cr & \null & + \frac{2}{3}\, g_{ij}(\phi) \, \partial_\mu \phi^i \,
\partial^\mu \phi^j  - {\cal V} (\phi) \,\Bigr )
\end{array}
\label{lagrared}
\end{equation}
The $28$ 1--forms $A^{\Lambda\Sigma}_\mu \, dx^\mu = -
A^{\Sigma\Lambda}_\mu \, dx^\mu $ transform in the $28$
antisymmetric representation of the electric subgroup
$SL(8,\IR) \,\subset \,  E_{7(7)}$.
We have labeled these
vector fields with a pair of antisymmetric indices
\footnote{For later convenience we have slightly
changed the conventions of \cite{noi3}}, each of them
ranging on $8$ values: $\Lambda,\Sigma,\Delta,\Pi, ...=1, \dots, 8$.
Other fields 
of $N=8$ supergravity are, besides the vierbein 1--form
$V^a$ (yielding the metric $g_{\mu \nu})$, the  gravitino 1-form
$\psi_A=\gamma_5\psi_A$ and  the dilatino 0--form
$\chi_{ABC} =\gamma_5 \chi_{ABC}$ (anti-symmetric in $ABC$).
The latin indices  carried by the fermionic fields also
range on $8$ values $A,B,C,..=1,\dots, 8$ but they transform
covariantly under the compensating subgroup $SU(8)\subset E_{7(7)} $
rather then under $SL(8,\IR)$. 
Furthermore we use the convention that
raising and lowering the $SU(8)$ indices also changes the chirality
projection of the fermion fields so that $\psi^A=-\gamma_5\psi^A$
and $\chi^{ABC} =-\gamma_5 \chi^{ABC}$. \par
The spectrum is completed by the $70$ scalars  $\phi^i$ that 
can be identified with any set of coordinates for the non--compact 
coset manifold $E_{7(7)}/SU(8)$. 
As emphasized in \cite{noi3,noi2,noi1}
the best parametrization of the scalar sector
is provided by the solvable coordinates, namely by the
parameters of the solvable group to which the non--compact coset
manifold is metrically equivalent. This has been shown to be the case
for the construction of BPS black--hole solutions of ungauged supergravity
and it  will   prove to be the case in
the analysis of the potentials for gauged supergravity models. Yet  at
the level of the present paper, whose aim is an exhaustive
classification of the possible gaugings \footnote {
  the analysis of the corresponding potentials is postponed  to
a future publication}, the choice of a coordinate system on $E_{7(7)}/SU(8)$
is irrelevant: we never need to specify it in our reasoning.
\par

The only object which we need to manipulate is the coset
representative $\IL(\phi)$ parametrizing the equivalence classes
of $E_{7(7)}/SU(8)$.
Just to fix ideas and avoiding the subtleties of
the solvable decomposition we can think of $\IL(\phi)$  as
the exponential of the $70$--dimensional coset $\IK$ in the orthogonal
decomposition:
\begin{equation}
E_{7(7)} = SU(8) \, \oplus \, \IK
\label{ortodeco}
\end{equation}
In practice this means that we can write:
\begin{equation}
\IL(\phi)=\exp \left( \matrix{  0               &  \phi^{EFGH} \cr
                        \phi_{ABCD}   & 0              } \right)
 =\left( \matrix{  u^{\Lambda\Xi}{}_{AB}  &  v^{\Lambda\Xi CD} \cr
                   v_{\Delta\Gamma AB}
&  u_{\Delta\Gamma}{}^{CD}  } \right)  \,
\label{cosetrep}
\end{equation}
where the $70$ parameters $\phi_{ABCD}$ satisfy the self--duality
condition \footnote{
Here we have used the notation,
$\phi^{ABCD} \equiv (\phi_{ABCD})^*$} :
\begin{equation}
 \phi_{ABCD}=\frac{1}{4!}\varepsilon_{ABCDEFGH}\phi^{EFGH}
 \label{selfdual1}
\end{equation}

As it is well known, the interaction structure  
of the theory is fully
encoded in the following geometrical data:
\begin{enumerate}
\item The symplectic geometry of the scalar coset manifold
$ E_{7(7)}/SU(8)$
\item The choice of the gauge group ${\cal G}_{gauge} \, \subset \,
SL(8,\IR) \, \subset \, E_{7(7)}$
\end{enumerate}
In this paper we focus on the second item of the this list.  
But before doing this we recollect some information on
$g_{ij}$, ${\cal N}_{\Lambda\Sigma,\Delta\Pi}$ and ${\cal V}(\phi)$,
that are determined by these two items.
\par
Let us first recall that $g_{ij}$ appearing
in the scalar field kinetic term of the lagrangian
\eqn{lagrared} is the unique $E_{7(7)}$ invariant metric 
on the scalar coset manifold.
\par 
The period matrix ${\cal N}_{\Lambda\Sigma,\Delta\Pi}$ 
has the following general expression holding true for all
symplectically embedded coset manifolds \cite{gaizum}:
\begin{equation}
{\cal N}_{\Lambda\Sigma,\Delta \Pi}=h \cdot f^{-1}
\label{gaiazuma}
\end{equation}
The complex $ 28 \times 28 $ matrices $ f,h$ are
defined by the $Usp(28,28)$ realization $\IL_{Usp} \left(\phi\right)$ of the
coset representative which is related to its $Sp(56,\IR)$ counterpart
$\IL_{Sp}(\phi)$ through a Cayley transformation, as displayed in the
following formula \cite{amicimiei}:
\begin{eqnarray}
\IL_{Usp} \left(\phi\right) &=& \frac{1}{\sqrt{2}}\left(\matrix{ f+ {\rm i}h & \bar f+ {\rm i}\bar h \cr f- {\rm i}h
& \bar f - {\rm i}\bar h  \cr } \right) \nonumber\\
& \equiv & {\cal C} \, \IL_{Sp}\left(\phi\right) {\cal C}^{-1} \nonumber\\
\IL_{Sp}(\phi) & \equiv & \exp \left[ \phi^i \, T_i \right ] \, = \,
\left(\matrix{ A(\phi) & B(\phi) \cr C(\phi) & D(\phi) \cr } \right) \nonumber\\
{\cal C}& \equiv & \frac{1}{\sqrt{2}} \, \left(\matrix{ \bfone & {\rm i} \, \bfone \cr
\bfone & - \, {\rm i} \, \bfone \cr } \right)
\label{cayleytra}
\end{eqnarray}
The coset representative $\IL$ as defined by \eqn{cosetrep}
is in the $Usp(28,28)$ representation. As explained in \cite{noi3}
there are actually four bases where the $ 56 \times 56 $ matrix
$\IL(\phi)$ can be written:
\begin{enumerate}
\item {The $SpD(56)$--basis}
\item {The $UspD(28,28)$--basis}
\item {The $SpY(56)$--basis}
\item {The $UspY(28,28)$--basis}
\end{enumerate}
corresponding to two cases where $\IL$ is symplectic
real ($SpD(56)$,$SpY(56)$) and two cases where it is
pseudo--unitary symplectic ($UspD(56)$,$UspY(56)$). This further
distinction in a pair of subcases corresponds
to choosing either a basis composed of weights or of Young tableaux.
By relying on \eqn{cosetrep} we have chosen to utilize the
$UspY(28,28)$--basis which is directly related to the $SU(8)$ indices
carried by the fundamental fields of supergravity.
However, for the description
of the gauge generators the Dynkin basis is more convenient.
We can optimize the advantages of both bases introducing a mixed one
where  the coset representative $\IL $ is
multiplied on the left by the constant matrix ${\cal S}$ performing the
transition from the pseudo--unitary Young basis to the real symplectic
Dynkin basis. Explicitly we have:
\begin{eqnarray}
\left( \matrix { u^{AB} \cr  v_{AB} \cr }\right) & = & {\cal S} \,
\left( \matrix { W^{i} \cr  W^{i+28} \cr }\right) (i,1,\dots\, 28) \nonumber\\
\end{eqnarray}
where
\begin{eqnarray}
{\cal S}&=& \left( \matrix{ {\bf S} & {\bf 0} \cr {\bf 0} & {\bf S^\star}
\cr } \right) \, {\cal C} = \frac{1}{\sqrt{2}}\left( \matrix{ {\bf S} & {\rm i}\,{\bf S} \cr
{\bf S^\star} & -{\rm i}\,{\bf S^\star} \cr } \right) \nonumber\\
\end{eqnarray}
the $ 28 \times 28 $ matrix ${\bf S}$ being unitary:
\begin{eqnarray}
&& \null \nonumber\\
 {\bf S}^\dagger {\bf S} &=& \bfone
\end{eqnarray}
The explicit form of the $U(28)$ matrix ${\bf S}$ was given in
section 5.4 of \cite{noi3} while the weights of the
$E_{7(7)}$ ${\bf 56}$ representation are listed in table \ref{e7weight}.
\begin{table}[ht]\caption{{\bf
Weights of the ${\bf 56}$ representation of $E_{7(7)} $:}}
\label{e7weight}
\begin{center}
\begin{tabular}{||cl|c|cl||}
\hline
\hline
    Weight & $q^\ell$  & \null & Weight & $q^\ell$ \\
      name & vector  & \null & name & vector \\
\hline
\null & \null & \null & \null & \null \\
$ {\vec W}^{(1)} \, =\, $ &$  \{ 2,3,4,5,3,3,1\} $  & \null &
$ {\vec W}^{(2)} \, =\, $ &$  \{ 2,2,2,2,1,1,1\} $   \\
$ {\vec W}^{(3)} \, =\, $ &$  \{ 1,2,2,2,1,1,1\} $  & \null &
$ {\vec W}^{(4)} \, =\, $ &$  \{ 1,1,2,2,1,1,1\} $    \\
$ {\vec W}^{(5)} \, =\, $ &$  \{ 1,1,1,2,1,1,1\} $  & \null &
$ {\vec W}^{(6)} \, =\, $ &$  \{ 1,1,1,1,1,1,1\} $    \\
$ {\vec W}^{(7)} \, =\, $ &$  \{ 2,3,3,3,1,2,1\} $  & \null &
$ {\vec W}^{(8)} \, =\, $ &$  \{ 2,2,3,3,1,2,1\} $    \\
$ {\vec W}^{(9)} \, =\, $ &$  \{ 2,2,2,3,1,2,1\} $  & \null &
$ {\vec W}^{(10)} \, =\, $ &$  \{ 2,2,2,2,1,2,1\} $   \\
$ {\vec W}^{(11)} \, =\, $ &$  \{ 1,2,2,2,1,2,1\} $  & \null &
$ {\vec W}^{(12)} \, =\, $ &$  \{ 1,1,2,2,1,2,1\} $    \\
$ {\vec W}^{(13)} \, =\, $ &$  \{ 1,1,1,2,1,2,1\} $  & \null &
$ {\vec W}^{(14)} \, =\, $ &$  \{ 1,2,2,3,1,2,1\} $    \\
$ {\vec W}^{(15)} \, =\, $ &$  \{ 1,2,3,3,1,2,1\} $  & \null &
$ {\vec W}^{(16)} \, =\, $ &$  \{ 1,1,2,3,1,2,1\} $    \\
$ {\vec W}^{(17)} \, =\, $ &$  \{ 2,2,2,2,1,1,0\} $  & \null &
$ {\vec W}^{(18)} \, =\, $ &$  \{ 1,2,2,2,1,1,0\} $    \\
$ {\vec W}^{(19)} \, =\, $ &$  \{ 1,1,2,2,1,1,0\} $  & \null &
$ {\vec W}^{(20)} \, =\, $ &$  \{ 1,1,1,2,1,1,0\} $    \\
$ {\vec W}^{(21)} \, =\, $ &$  \{ 1,1,1,1,1,1,0\} $  & \null &
$ {\vec W}^{(22)} \, =\, $ &$  \{ 1,1,1,1,1,0,0\} $    \\
$ {\vec W}^{(23)} \, =\, $ &$  \{ 3,4,5,6,3,4,2\} $  & \null &
$ {\vec W}^{(24)} \, =\, $ &$  \{ 2,4,5,6,3,4,2\} $    \\
$ {\vec W}^{(25)} \, =\, $ &$  \{ 2,3,5,6,3,4,2\} $  & \null &
$ {\vec W}^{(26)} \, =\, $ &$  \{ 2,3,4,6,3,4,2\} $    \\
$ {\vec W}^{(27)} \, =\, $ &$  \{ 2,3,4,5,3,4,2\} $  & \null &
$ {\vec W}^{(28)} \, =\, $ &$  \{ 2,3,4,5,3,3,2\} $    \\
$ {\vec W}^{(29)} \, =\, $ &$  \{ 1,1,1,1,0,1,1\} $  & \null &
$ {\vec W}^{(30)} \, =\, $ &$  \{ 1,2,3,4,2,3,1\} $    \\
$ {\vec W}^{(31)} \, =\, $ &$  \{ 2,2,3,4,2,3,1\} $  & \null &
$ {\vec W}^{(32)} \, =\, $ &$  \{ 2,3,3,4,2,3,1\} $    \\
$ {\vec W}^{(33)} \, =\, $ &$  \{ 2,3,4,4,2,3,1\} $  & \null &
$ {\vec W}^{(34)} \, =\, $ &$  \{ 2,3,4,5,2,3,1\} $    \\
$ {\vec W}^{(35)} \, =\, $ &$  \{ 1,1,2,3,2,2,1\} $  & \null &
$ {\vec W}^{(36)} \, =\, $ &$  \{ 1,2,2,3,2,2,1\} $   \\
$ {\vec W}^{(37)} \, =\, $ &$  \{ 1,2,3,3,2,2,1\} $  & \null &
$ {\vec W}^{(38)} \, =\, $ &$  \{ 1,2,3,4,2,2,1\} $    \\
$ {\vec W}^{(39)} \, =\, $ &$  \{ 2,2,3,4,2,2,1\} $  & \null &
$ {\vec W}^{(40)} \, =\, $ &$  \{ 2,3,3,4,2,2,1\} $   \\
$ {\vec W}^{(41)} \, =\, $ &$  \{ 2,3,4,4,2,2,1\} $  & \null &
$ {\vec W}^{(42)} \, =\, $ &$  \{ 2,2,3,3,2,2,1\} $    \\
$ {\vec W}^{(43)} \, =\, $ &$  \{ 2,2,2,3,2,2,1\} $  & \null &
$ {\vec W}^{(44)} \, =\, $ &$  \{ 2,3,3,3,2,2,1\} $    \\
$ {\vec W}^{(45)} \, =\, $ &$  \{ 1,2,3,4,2,3,2\} $  & \null &
$ {\vec W}^{(46)} \, =\, $ &$  \{ 2,2,3,4,2,3,2\} $    \\
$ {\vec W}^{(47)} \, =\, $ &$  \{ 2,3,3,4,2,3,2\} $  & \null &
$ {\vec W}^{(48)} \, =\, $ &$  \{ 2,3,4,4,2,3,2\} $    \\
$ {\vec W}^{(49)} \, =\, $ &$  \{ 2,3,4,5,2,3,2\} $  & \null &
$ {\vec W}^{(50)} \, =\, $ &$  \{ 2,3,4,5,2,4,2\} $    \\
$ {\vec W}^{(51)} \, =\, $ &$  \{ 0,0,0,0,0,0,0\} $  & \null &
$ {\vec W}^{(52)} \, =\, $ &$  \{ 1,0,0,0,0,0,0\} $    \\
$ {\vec W}^{(53)} \, =\, $ &$  \{ 1,1,0,0,0,0,0\} $  & \null &
$ {\vec W}^{(54)} \, =\, $ &$  \{ 1,1,1,0,0,0,0\} $   \\
$ {\vec W}^{(55)} \, =\, $ &$  \{ 1,1,1,1,0,0,0\} $  & \null &
$ {\vec W}^{(56)} \, =\, $ &$  \{ 1,1,1,1,0,1,0\} $   \\
\hline
\end{tabular}
\end{center}
\end{table}
\subsection{The supersymmetry transformation rules}
To complete our illustration of the bosonic lagrangian we 
discuss the scalar potential ${\cal V}(\phi)$. This cannot be done
without referring to the supersymmetry transformation rules because
of its crucial relation with the so called {\it fermion shifts} that appear
in such rules and that are the primary objects determined by the
choice of the gauge algebra.
\par
Since the $N=8$ theory has no matter multiplets the
fermions are just, as already remarked,
the ${\bf 8}$ spin $3/2$ gravitinos and the
${\bf 56}$ spin $1/2$ dilatinos. The two numbers ${\bf 8}$ and ${\bf
56}$ have been written boldfaced since they also single out the
dimensions of the two irreducible $SU(8)$ representations to which
the two kind of fermions are respectively assigned, namely the
fundamental and the three times antisymmetric:
\begin{equation}
 \psi_{\mu\vert A} \, \leftrightarrow \,
 \mbox{$ \begin{array}{|c|}
\hline
\stackrel{ }{A}\cr
\hline
\end{array}$} \, \equiv \,  {\bf 8}\quad ; \quad \chi _{ABC}  \,
\leftrightarrow \,
\mbox{$ \begin{array}{|c|}
\hline
\stackrel{ }{A}\cr
\hline
\stackrel{}{B}\cr
\hline
\stackrel{}{C}\cr
\hline
\end{array}
$}   \, \equiv \,  {\bf 56}
\end{equation}
Following the conventions and formalism of \cite{castdauriafre},
\cite{amicimiei}, \cite{noi3} the fermionic  supersymmetry
transformation rules of are written as follows:
\begin{eqnarray}
\delta \psi _{A\mu }&=&\nabla_\mu \epsilon_A + \,\frac{\sqrt{3}}{4 \,
\sqrt{2}}
T^{-}_{AB\vert \rho\sigma} \,\gamma^{\rho\sigma} \, \gamma_\mu  \,
\epsilon^B  +  S_{AB} \, \gamma_\mu  \, \epsilon^B \, +
\cdots \nonumber\\
\delta \chi_{ABC} &=& a P_{ABCD\vert i } \, \partial_\mu \phi^i
\, \gamma^\mu  \, \epsilon^D + b \,
T^{-}_{[AB \vert \rho\sigma} \,\gamma^{\rho\sigma}
\epsilon_{C ]}  +  \Sigma^{D}_{ABC} \, \epsilon_D \cdots
\label{trasforma}
\end{eqnarray}
where $a,b$ are numerical coefficients fixed
by superspace  Bianchi identities,
$T^{-}_{AB\vert \mu\nu}$ is the antiselfdual part of the
graviphoton field strength, $P_{ABCD\vert i }$ is the vielbein of the scalar
coset manifold completely antisymmetric in $ABCD$ and satisfying the
same pseudoreality condition as our choice of the scalars $\phi_{ABCD}$:
\begin{equation}
P_{ABCD}=\frac{1}{4!}\epsilon_{ABCDEFGH}\bar P^{EFGH}.
\end{equation}
and $S_{AB}$,$\Sigma^{D}_{ABC}$ are the {\it fermion shifts} that, as
we will see later, determine the potential ${\cal V}(\phi)$.
\par
What we need is the explicit expression of these objects in terms
of the coset representatives.
For the graviphoton such an expression is {\it gauging independent} and
coincides with that appearing in the case of ungauged supergravity.
On the other hand, for the scalar vielbein and the fermion shifts,
their expression involves the choice of the gauge group and can be given
only upon introduction of the {\it gauged  Maurer Cartan equations}.
Hence we first   recall  the structure of the graviphoton and
then we turn our attention to the second kind of items entering the
transformation rules that are the most relevant ones for our
discussion.
\subsubsection{The graviphoton field strength}
We  introduce the multiplet of electric and magnetic field strengths
according to the standard definitions of \cite{n2paperone},\cite{mylecture}
\cite{ricsertoi}:
\begin{equation}
{\vec V}_{\mu\nu} \equiv \left(\matrix {
F^{\Lambda\Sigma}_{\mu\nu} \cr G_{\Delta\Pi\vert\mu\nu} \cr}\right)
\label{symvecft}
\end{equation}
where
\begin{eqnarray}
G_{\Delta\Pi\vert\mu\nu} &=& -\mbox{Im}{\cal N}_{\Delta\Pi,\Lambda\Sigma} \,
{\widetilde F}^{\Lambda\Sigma}_{\mu\nu} -
\mbox{Re}{\cal N}_{\Delta\Pi,\Lambda\Sigma} \,
{ F}^{\Lambda\Sigma}_{\mu\nu}\nonumber\\
{\widetilde F}^{\Lambda\Sigma}_{\mu\nu}&=&\frac{1}{2}\, \epsilon_{\mu
\nu\rho\sigma} \, F^{\Lambda\Sigma\vert \rho\sigma}
\end{eqnarray}
The $56$--component field strength vector ${\vec V}_{\mu\nu}$
transforms in the real symplectic representation of the U--duality
group $E_{7(7)}$. We can also write  a column vector containing
the $ 28 $ components of the graviphoton field strengths and their
complex conjugate:
\begin{equation}
{\vec T}_{\mu\nu} \equiv \left(\matrix{
T^{\phantom{\mu \nu}\vert AB}_{\mu \nu}  \cr
T_{\mu \nu \vert AB} \cr }\right) \quad T^{\vert AB}_{\mu \nu} =
\left(T_{\mu \nu \vert AB}\right)^\star
\label{gravphotvec}
\end{equation}
in which the upper and lower components  transform in the canonical
{\it Young basis} of $SU(8)$ for the ${\bar {\bf 28}}$ and  ${\bf 28}$
representation respectively.\par
The relation between the graviphoton field strength vectors and the
electric magnetic field strength vectors involves the coset
representative in the $SpY(56)$ representation and it is the following one:
\begin{equation}
{\vec T}_{\mu \nu} = - {\cal C} \, \IC \, \IL_{SpY}^{-1}(\phi) \, {\vec
V}_{\mu \nu}
\label{T=SCLV}
\end{equation}
The matrix
\begin{equation}
\IC =\left(\matrix { {\bf 0} & \bfone \cr -\bfone & {\bf 0} }\right)
\label{sympinv}
\end{equation}
is the symplectic invariant matrix. Eq.\eqn{T=SCLV} reveal that the
graviphotons transform under the $SU(8)$ compensators associated
with the $E_{7(7)}$ rotations.
It is appropriate to express the upper and lower components of $\vec{T}$
in terms of the self--dual and antiself--dual parts of the graviphoton
field strengths, since only the latters enter (\ref{trasforma}) \\
These components are defined as follows:
\begin{eqnarray}
T^{+\vert AB}_{ \mu \nu}&=& \frac{1}{2}\left(T^{ \vert AB}_{\mu \nu}+
\frac{{\rm i}}{2}
\, \epsilon_{\mu\nu\rho\sigma}g^{\rho\lambda} g^{\sigma\pi}
\, T^{ \vert AB}_{\lambda \pi}\right) \nonumber\\
T^-_{ AB \vert \mu \nu}&=& \frac{1}{2}\left(T_{  AB \vert\mu \nu}-\frac{{\rm i}}{2}
\, \epsilon_{\mu\nu\rho\sigma}g^{\rho\lambda} g^{\sigma\pi}
\, T_{ AB \vert\lambda \pi} \right)
\label{selfdual}
\end{eqnarray}
As shown in \cite{noi3} the following equalities hold true:
\begin{eqnarray}
 T_{\mu  \nu}^{\phantom{\mu  \nu}\vert AB}&=& T^{+ \vert AB}_{\mu  \nu} \nonumber\\
 T_{ \mu  \nu\vert AB}&=& T^{-}_{ \mu  \nu\vert AB}
\label{symprop}
\end{eqnarray}
and we can simply write:
\begin{equation}
{\vec T}_{\mu \nu} \equiv \left(\matrix{T^{+\vert AB}_{\mu \nu}  \cr
T^{-}_{\mu \nu \vert AB} \cr }\right)
\end{equation}
\subsubsection{The gauged Maurer Cartan equations and the fermion shifts}
As it is well known (see for instance the lectures \cite{mylecture})
the key ingredient in the gauging of an extended supergravity theory is
provided by the {\it gauged} left-invariant 1--forms on the coset manifold.
This notion is applied to the present case in the following way.
\par
First note that in the $UspY(28,28)$ basis we have chosen the coset
representative \eqn{cosetrep} satisfies the following identities:
\begin{eqnarray}
u^{\Pi\Delta}{}_{AB}u^{AB}{}_{\Lambda\Sigma}-
v^{\Pi\Delta AB}v_{AB\Lambda\Sigma}&=&
\delta^{\Pi\Delta}_{\Lambda\Sigma}\,,\nonumber\\
u^{\Pi\Delta}{}_{AB}v^{AB\Lambda\Sigma}-
v^{\Pi\Delta AB}u_{AB}{}^{\Lambda\Sigma}&=&0\,,
\label{identuv}
\end{eqnarray}
\begin{eqnarray}
u^{AB}{}_{\Lambda\Sigma}u^{\Lambda\Sigma}{}_{CD}-
v^{AB \Lambda\Sigma}v_{\Lambda\Sigma CD}&=&
\delta^{AB}_{CD}\,,\nonumber\\
u^{AB}{}_{\Lambda\Sigma}v^{\Lambda\Sigma CD}-
v^{AB\Lambda\Sigma}u_{\Lambda\Sigma}{}^{CD}&=&0\,,
\label{2identuv}
\end{eqnarray}
and the inverse coset representative is given by:
\begin{equation}
\IL^{-1} =\left( \matrix{  u^{AB}{}_{\Lambda\Sigma}  &
-v^{AB\Delta\Gamma} \cr
                   -v_{CD\Lambda\Sigma}    &  u_{CD}{}^{\Delta\Gamma}  }
                   \right)  \,.
\label{Linvers}
\end{equation}
where, by raising and lowering the indices, complex conjugation
is understood.
\par
Secondly recall that in our basis the generators of the electric
subalgebra $SL(8,\IR) \, \subset \, E_{7(7)}$ have the following
form
\begin{equation}
G_\alpha  =
\left( \matrix{    q^{\Lambda\Sigma}{}_{\Pi\Delta}(\alpha )  &
                    p^{\Lambda\Sigma\Psi\Xi} (\alpha ) \cr
                   p_{\Delta\Gamma\Pi\Delta}(\alpha )  &
                   q_{\Lambda\Sigma}{}^{\Psi\Xi}(\alpha )  } \right)
\label{sl8gener}
\end{equation}
where the matrices $q$ and $p$ are real and have the following form
\begin{eqnarray}
q^{\Lambda\Sigma}{}_{\Pi\Delta}&=&2\delta^{[\Lambda}{}_{[\Pi}q^{\Sigma]}
{}_{\Delta]}={2\over 3}
\delta^{[\Lambda}{}_{[\Pi}q^{\Sigma]\Gamma}{}_{\Delta]\Gamma}\,,\nonumber\\
p_{\Delta\Gamma\Pi\Omega}&=&{1\over 24}
\varepsilon_{\Delta\Gamma\Pi\Omega\Lambda\Sigma\Psi\Xi}
p^{\Lambda\Sigma\Psi\Xi}\,.\label{uspconstraints}
\end{eqnarray}
The index $\alpha=1,\dots \, 63$  in (\ref{sl8gener}) spans the adjoint 
representation of
$SL(8,\IR)$ according to some chosen basis and we can freely raise
and lower the greek indices $ \Lambda, \Sigma, ...$ because of the
reality of the representation.
\par
Next let us introduce the fundamental item in the gauging
construction. It is the $28 \times 63 $ constant embedding matrix:
\begin{equation}
{\bf {\cal E}} \equiv  e_{\Lambda\Sigma}^\alpha
\label{alettomat}
\end{equation}
transforming under $SL(8,\IR)$ as its indices specify, namely in the
tensor product of the adjoint with the antisymmetric ${\bf 28}$ and
that specifies which
generators of $SL(8,\IR)$ are gauged and by means of which vector
fields in the $28$--dimensional stock. In particular, using this matrix
${\cal E}$, one write the gauge 1--form as:
\begin{equation}
 A \equiv A^{\Lambda\Sigma} e_{\Lambda\Sigma}^\alpha G_\alpha
 \label{gaugeform}
\end{equation}
The main result in the present paper will be the determination of the
most general form  and the analysis of the embedding matrix  
$e_{\Lambda\Sigma}^\alpha$.
In terms of the gauge 1--form $A$ and of the coset representative
$\IL(\phi)$ we can write the {\it gauged left--invariant 1--form}:
\begin{equation}
\Omega=\IL^{-1} d L + g \IL^{-1} A \IL \,
\label{gaul1f}
\end{equation}
which   belongs to the $E_{7(7)}$ Lie algebra in the
$UspY(28,28)$ representation and defines the {\it gauged} scalar
vielbein $P^{ABEF}$ and the $SU(8)$ connection $Q^{\phantom{D} B}_{D}$:
\begin{equation}
\Omega = \, \left( \matrix{
2 \delta^{[A }_{[C} \, Q^{\phantom{D} B]}_{D]}
& P^{ABEF} \cr P_{CDGH} &
- 2 \delta^{[E}_{[G} \, Q^{ F]}_{\phantom{F}H]}\cr }\right)
\label{uspYconnec}
\end{equation}
Because of its definition  the 1--form $\Omega$ satisfies
{\it gauged Maurer Cartan equations}:
\begin{equation}
d\Omega + \Omega\wedge\Omega =
g \big[F^{\Lambda\Sigma}
-\big( \sqrt{2}(u^{\Lambda\Sigma}{}_{AB}+v_{\Lambda\Sigma AB})
\bar\psi^A \wedge \psi^B + {\rm h.c.} \big)\big]
e_{\Lambda\Sigma}^\alpha \IL^{-1} G_\alpha  \IL\,,
\label{MaurerCartan}
\end{equation}
with $F^{\Lambda\Sigma} $ the supercovariant field strength of the vectors
$A^{\Lambda\Sigma}$. Let us focus on the last factor in
eq.\eqn{MaurerCartan}:
\begin{equation}
{\bf U}_\alpha  \, \equiv \, \IL^{-1} G_\alpha  \IL  =
\left( \matrix{    \cal A (\alpha ) &  \cal B (\alpha )\cr
                   \bar{\cal B} (\alpha )  &  \bar{\cal A} (\alpha ) } \right)
\,
\label{boosted}
\end{equation}
Since ${\bf U}_\alpha$ is an $E_{7(7)}$ Lie algebra element,
for each gauge generator $G_\alpha $ we necessarily have:
\begin{eqnarray}
{\cal A}^{AB}_{\phantom{AB}CD}(\alpha ) &=& \frac{2}{3} \,
\delta^{[A }_{[C} \, {\cal A}^{B]M}_{\phantom{B]M}D]M} \nonumber\\
 {\cal B}^{ABFG}(\alpha ) &=& {\cal B}^{[ABFG]}(\alpha )
\label{proprieta}
\end{eqnarray}
Comparing with eq.\eqn{MaurerCartan} we see that the scalar field
dependent $SU(8)$ tensors multiplying the gravitino bilinear terms
are the following ones:
\begin{eqnarray}
T^A{}_{BCD}& \equiv & (u^{\Omega\Sigma}{}_{CD} + v_{\Omega\Sigma CD}) \,
              e^\alpha_{\Omega\Sigma} \, {\cal A}^{AM}{}_{BM}(\alpha)
\label{Ttensor} \nonumber\\
Z_{CD}^{ABEF}&  \equiv &
(u^{\Omega\Sigma}{}_{CD} + v_{\Omega\Sigma CD}) \,
              e^\alpha_{\Omega\Sigma} \, {\cal B}^{ABEF}(\alpha )
\label{uvetta}
\end{eqnarray}
As shown in the original papers by de Wit and Nicolai \cite{dwni}
(or Hull \cite{hull}) and reviewed in \cite{castdauriafre},
closure of the supersymmetry algebra
\footnote{ in the rheonomy approach closure of
the Bianchi identities} and hence existence of the corresponding
gauged supergravity models is obtained {\it if and only if}
the following $T$--identities are satisfied:
\begin{eqnarray}
T^A{}_{BCD} &=& T^A{}_{[BCD]} + \frac{2}{7} \delta^A_{[C} T^M{}_{D]MB}
\label{T-id} \\
Z^{CD}_{ABEF}&=&\frac{4}{3} \delta^{[C}{}_{[A} T^{D]}{}_{BEF]}
\label{T-idbis}
\end{eqnarray}
Eq.s \eqn{T-id} and \eqn{T-idbis} have a clear group theoretical
meaning. Namely, they state that both the $T^A{}_{BCD}$ tensor and
the $Z_{CD}^{ABEF}$ tensor can be expressed in basis spanned by two
irreducible $SU(8)$ tensors corresponding to the {\bf
420} and {\bf 36} representations  respectively:
\begin{equation}
{\stackrel {\circ}{^{\phantom{A}}T^A}}_{BCD}
\,  \equiv \,  \epsilon^{AI_1\dots I_7} \,
\mbox{$ \begin{array}{l}
{\begin{array}{|c|c|}
\hline
\stackrel{ }{I_1} & \stackrel { }{B}\cr
\hline
\stackrel{ }{I_2} & \stackrel { }{C}\cr
\hline
\stackrel{ }{I_3} & \stackrel { }{D}\cr
\hline
\end{array} } \cr
{\begin{array}{|c|}
\hline
\stackrel{ }{I_4} \cr
\hline
\end{array} }  \cr
{\begin{array}{|c|}
\hline
\stackrel{ }{I_5} \cr
\hline
\end{array} } \cr
{\begin{array}{|c|}
\hline
\stackrel{ }{I_6} \cr
\hline
\end{array} }   \cr
{\begin{array}{|c|}
\hline
\stackrel{ }{I_7} \cr
\hline
\end{array} } \cr
\end{array}$ }  \, \equiv \, {\bf 420} \nonumber  \, \qquad ; \qquad
 {\stackrel {\circ}{T}}_{DB}  \,  \equiv \,  \mbox{$
 \begin{array}{|c|c|}
\hline
\stackrel{ }{D} & \stackrel { }{B}\cr
\hline
\end{array} $} \, \equiv \, {\bf 36}
\label{representazie}
\end{equation}
 To see this let us consider first eq. \eqn{T-id}. In general a
 tensor of type $T^A{}_{B[CD]}$  would have $8 \times 8 \times 28$
 components and contain several irreducible representations of $SU(8)$. However, as a
 consequence of eq. \eqn{T-id} only the representations {\bf 420},
 {\bf 28} and {\bf 36} can appear. (see fig. $1$).
 In addition, since the $\cal A$ tensor, being in the adjoint of $SU(8)$, is traceless
 also the $T$-tensor
appearing in \eqn{T-id} is traceless: $T^A{}_{ABC}=0$. Combining this information   with
eq.\eqn{T-id}  we obtain
\begin{equation}
T^M{}_{[AB]M}=0,
\label{no28}
\end{equation}
Eq. \eqn{no28} is the statement that the {\bf 28} representation
appearing in fig. $1$ vanishes so that the $T^A{}_{B[CD]}$
tensor is indeed expressed solely in terms of the irreducible tensors
\eqn{representazie}.\par
\vskip 0.5cm
\begin{figure}[ht]
\label{young1}
\caption{Decomposition in irreducible representations of a tensor of type $T^A{}_{BCD}$  }
\centerline{\epsfig{figure=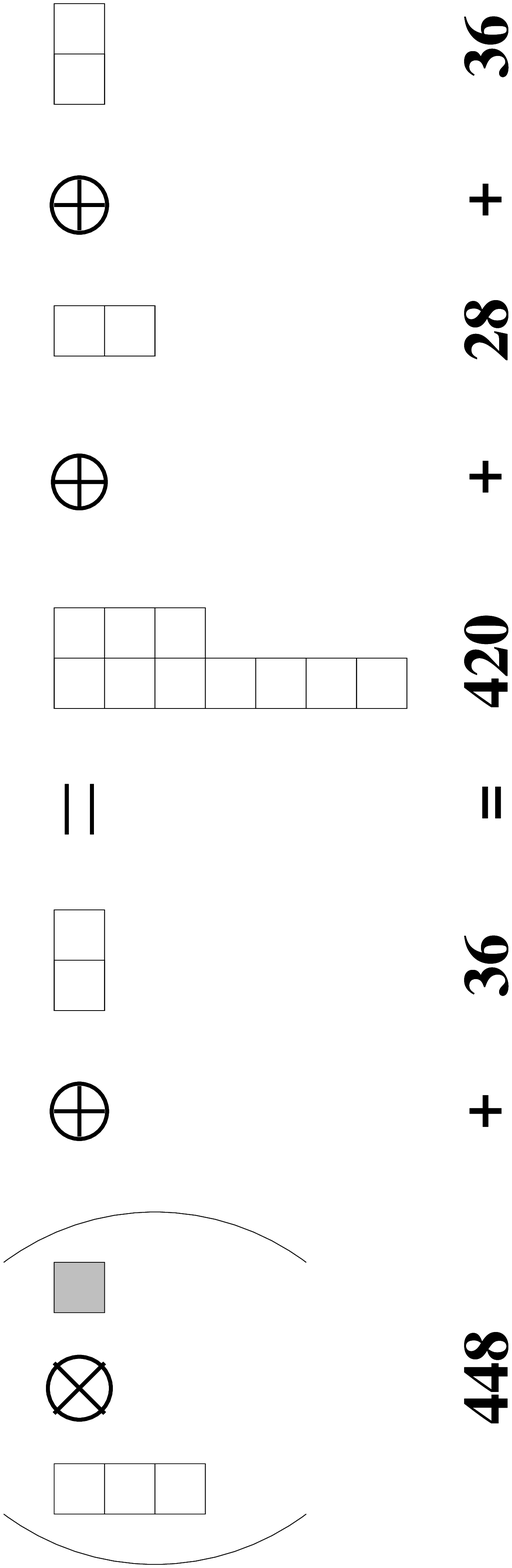,height=11cm,width=5cm,angle=-90}}
\end{figure}
\bigskip
\noindent
A similar argument can be given to interpret the second $T$--identity \eqn{T-idbis}.
A tensor of type $Z_{[CD]}^{[ABEF]}$ contains, a priori, $70 \times 28$
components and contains the irreducible representations  {\bf 1512},
{\bf 420} and {\bf 28} (see fig. 2). Using eq.\eqn{T-idbis} one immediately
sees that the representations {\bf 1512} and {\bf 28} must vanish and
that the surviving  {\bf 420} is proportional through a fixed coefficient to the
{\bf 420} representations appearing in the decomposition of the $T^A{}_{B[CD]}$ tensor.
\begin{figure}[ht]
\label{young2}
\caption{Decomposition in irreducible representations of a tensor of type $Z^{CD}_{ABEF}$ }
\centerline{\epsfig{figure=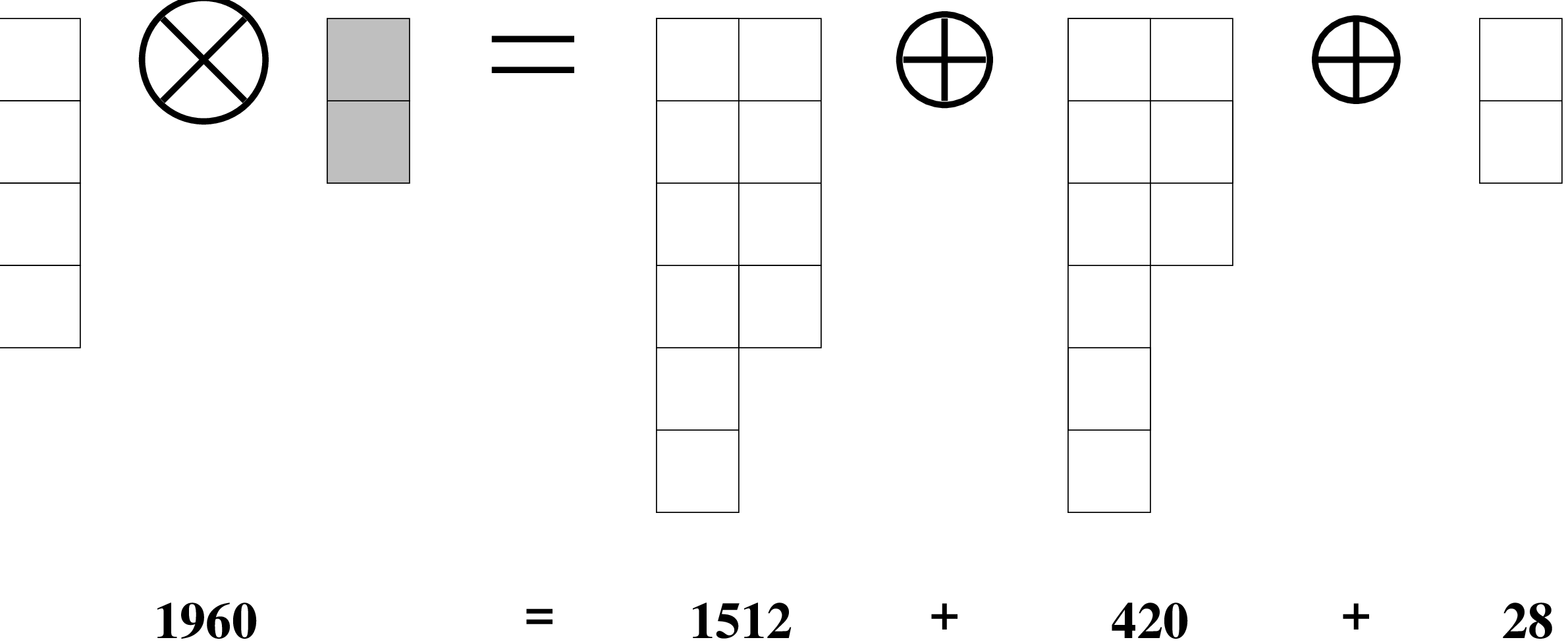,height=7cm,width=3.5cm,angle=-90}}
\end{figure}
\bigskip
\noindent
In view of this discussion, the $T$--identities can be rewritten as
follows in the basis of the independent irreducible tensors
\begin{eqnarray}
{\stackrel {\circ}{^{\phantom{A}}T^A}}{}_{ BCD } &=& T^A{}_{[BCD]}
 \nonumber\\
{\stackrel {\circ}{T}}_{AB} &=& T^M{}_{AMB} 
\label{panettone}
\end{eqnarray}
The irreducible tensors {\bf 420} and {\bf 36}   can be identified, through a
suitable coefficient fixed by Bianchi identities, with the fermion shifts appearing
in the supersymmetry transformation rules \eqn{trasforma}:
\begin{equation}
\Sigma^{A}_{BCD} \, = \, \sigma  {\stackrel {\circ}{^{\phantom{A}}T^A}}_{ BCD } \quad ; \quad
S_{DB} \, = \, s \, {\stackrel {\circ}{T}}_{DB}
\label{agnizio}
\end{equation}
\subsubsection{The Ward identity and the potential}
Finally, according to a general mechanism present in all extended
supergravities \cite{Wardide} the scalar potential is defined
by the following supersymmetry Ward identity:
\begin{equation}
-{\cal V} \,\delta^{A}_{B} \,  = \, S_{AM} \, S^{BM} \, - \,
\Sigma_{A}^{PQR} \, \Sigma^{B}_{PQR}
\end{equation}
which, as shown by de Wit and Nicolai \cite{dwni} is satisfied if
and only if the ratio between the two constants in eq. \eqn{agnizio}
is:
\begin{equation}
 \frac{s^2}{\sigma^2} \, = \, \frac{18}{49}
 \label{razione}
\end{equation}
\section{Algebraic characterization of the gauge group embedding
$G_{gauge} \longrightarrow SL(8,\IR)$}
As we have seen in the previous section the existence of gauged
supergravity models relies on a peculiar pair of identities to
be satisfied by the $T$--tensors. Therefore a classification of all
possible gaugings involves a parametrization of all $SL(8,\IR)$
subalgebras that lead to satisfied $T$--identities. Since the
$T$--tensors are scalar field dependent objects it is not immediately
obvious how such a program can be carried through. On the other
hand since the problem is algebraic in nature (one looks for all
Lie subalgebras of $SL(8,\IR)$ fulfilling a certain property) it is
clear that it should admit a completely algebraic formulation.
It turns out that such an algebraic formulation is possible
and actually very simple.  We will show that the $T$--identities
imposed on the $T$--tensors are nothing else but a single algebraic
equation imposed on the embedding matrix ${\cal E}$
introduced in eq.\eqn{alettomat}. This is what we do in the present
section.
\par
To begin with we recall a general and obvious constraint to be satisfied
by ${\cal E}$ which embeds a subalgebra of the $SL(8,\IR)$ Lie
algebra into its ${\bf 28}$ irreducible representation.
As it was already stressed in \cite{noi2}, the vectors should be in the
coadjoint representation of the gauge group. Hence under
reduction to $G_{gauge}\subset SL(8,\IR)$ we must obtain the following
decomposition of the entire set of the electric vectors:
\begin{equation}
{\bf 28}\stackrel{G_{gauge}}{\rightarrow }{\bf coadj} {G_{gauge}}
\oplus {\cal R}
\label{adjdec}
\end{equation}
where ${\cal R}$ denotes the subspace of vectors not entering the
adjoint representation of $G_{gauge}$ which is not necessarily a
representation of $G_{gauge}$ itself.
\par
Next in order to reduce the field dependent $T$--identities to an
algebraic equation on ${\cal E}$ we introduce the following constant
tensors:\footnote{
For example, in the de Wit--Nicolai theory, where one gauges
$G_{gauge} =SO(8)$ we have:
\[
t^{(1)}_{\Omega\Sigma}{}^{\Pi\Gamma}{}_{\Delta\Lambda}=
\delta_{[\Delta}^{[\Pi}\delta_{\Lambda][\Omega}
\delta_{\Sigma]}^{\Gamma]}\,,\qquad t^{(2)}_{\Omega\Sigma}{}^{\Pi\Gamma\Delta\Lambda}=0  \,.
\]
}
\begin{eqnarray}
 t^{(1)}_{\Omega\Sigma}{}^{\Pi\Gamma}{}_{\Delta\Lambda}
&\equiv& \sum_{\alpha } e^\alpha_{\Omega\Sigma} \,
q^{\Pi\Gamma}{}_{\Delta\Lambda}(\alpha ) \,,
\, \nonumber\\
t^{(2)}_{\Omega\Sigma}{}^{\Pi\Gamma\Delta\Lambda}
&\equiv& \sum_{\alpha } e^\alpha_{\Omega\Sigma} \,
p^{\Pi\Gamma\Delta\Lambda}(\alpha )
\,.
\end{eqnarray}
In terms of $t^{(1)}$ and  $t^{(2)}$  the field dependent $T$-tensor is
rewritten as
\begin{eqnarray}
T^A{}_{BCD} = (u^{\Omega\Sigma}{}_{CD} + v_{\Omega\Sigma CD})
             \big[\,
             t^{(1)}_{\Omega\Sigma}{}^{\Pi\Gamma}{}_{\Delta\Lambda}\,
                  (u^{AM}{}_{\Pi\Gamma} \, u^{\Delta\Lambda}{}_{BM}
                   - v^{AM\Phi\Gamma} \, v_{\Delta\Lambda BM} ) \,\,
                   \nonumber\\
                  + t^{(2)}_{\Omega\Sigma}{}^{\Pi\Gamma\Delta\Lambda}\,
                  (u^{AM}{}_{\Pi\Gamma} \, v^{\Delta\Lambda}{}_{BM}
                   - v^{AM\Phi\Gamma} \, u_{\Delta\Lambda BM} ) \, \big]\,.
\end{eqnarray}
Then we can state our main result as the following
\bth
{\large The field dependent $T$-identities are fully equivalent to the
following algebraic equation:}
\begin{eqnarray}
t^{(1)}_{\Omega\Sigma}{}^{\Pi\Gamma}{}_{\Delta\Lambda} + t^{(1)}_{\Delta\Lambda}{}^{\Pi\Gamma}{}_{\Omega\Sigma}
+ t^{(2)}_{\Pi\Gamma}{}^{\Delta\Lambda}{}^{\Omega\Sigma} &=& 0
\,,  \nonumber
\\
\label{lt-id}
\end{eqnarray}
\eth
\bpr  {-}
\epr
{\small
We have to show that the field dependent $T$-identities
(\ref{T-id}) and (\ref{T-idbis}) are fully equivalent to the
algebraic expression \eqn{lt-id}. So we begin with one direction of
the proof
\begin{enumerate}
\item {{\sl   Proof that \eqn{lt-id}  implies \eqn{T-id}  and
\eqn{T-idbis} } \par
Summing three times (\ref{lt-id}) with permuted indices,  one finds
\begin{equation}
t^{(2)}_{\Omega\Sigma}{}^{\Pi\Gamma}{}^{\Delta\Lambda} + t^{(2)}_{\Delta\Lambda}{}^{\Pi\Gamma}{}^{\Omega\Sigma}
- t^{(2)}_{\Pi\Gamma}{}^{\Delta\Lambda}{}^{\Omega\Sigma} = 0  \,.
\label{lt-id2}
\end{equation}
>From eq.s (\ref{lt-id}), (\ref{lt-id2}) one finds that
\begin{equation}
(u^{AB}{}_{\Omega\Sigma} + v^{AB\Omega\Sigma})
e_{\Omega\Sigma}^\alpha {\cal B}_{CDEF}=
                {2\over 3}\delta^{[A}{}_{[B} T^{C]}{}_{D]GH}
               +{2\over 3}\delta^{[A}{}_{[G} T^{C]}{}_{H]BD}
\label{uveB}
\end{equation}
Since $ {\cal B}_{CDEF}$ is antisymmetric in $CDEF$,
one has
\begin{eqnarray}
(u^{AB}{}_{\Omega\Sigma} + v^{AB\Omega\Sigma})
e_{\Omega\Sigma}^\alpha {\cal B}_{CDEF}&=&
\frac{4}{3} \delta^{[A}{}_{[B} T^{C]}{}_{DGH]} \,,\nonumber \\
{2\over 3}\delta^{[A}{}_{[B} T^{C]}{}_{D]GH}
+{2\over 3}\delta^{[A}{}_{[G} T^{C]}{}_{H]BD}&=&
\frac{4}{3} \delta^{[A}{}_{[B} T^{C]}{}_{DGH]} \,.
\end{eqnarray}
The first of the above equations
coincides with eq. (\ref{T-idbis}). From the second one,   taking
suitable contractions, as
explained in \cite{dwni}, one obtains eq. (\ref{T-id}). This shows that
(\ref{lt-id}) implies both eq.s \eqn{T-id} and \eqn{T-idbis}.}
\item{{\sl Proof that (\ref{T-id}) and (\ref{T-idbis}) imply (\ref{lt-id}) }
\par
If we use (\ref{T-id}) to rewrite the following expression
\begin{eqnarray}
&&{2\over 3}\delta^{[A}{}_{[B} T^{C]}{}_{D]GH}
+{2\over 3}\delta^{[A}{}_{[G} T^{C]}{}_{H]BD}=\nonumber \\
&&=\left[{2\over 3}\delta^{[A}{}_{[B} T^{C]}{}_{D]GH}\right]_{\left[DGH\right]}
+\left[{2\over 3}\delta^{[A}{}_{[G} T^{C]}{}_{H]BD}\right]_{\left[HBD\right]}
+\nonumber \\
&&+\left[{4\over 21}\delta^{[A}{}_{[B}\delta^{[C}{}_{|G}
 T^{M}{}_{HM|D]}\right]_{\left[GH\right]}+
\left[{4\over 21}\delta^{[A}{}_{[G}\delta^{[C}{}_{|B}
 T^{M}{}_{DM|H]}\right]_{\left[BD\right]}
\end{eqnarray}
we can easily verify that it is antisymmetric in the indices
$GDBH$. Indeed  the last two terms cancel, while the sign in front of the first two terms is reversed
upon interchanging $B$ and $G$. This means that
\begin{equation}
{2\over 3}\delta^{[A}{}_{[B} T^{C]}{}_{D]GH}
+{2\over 3}\delta^{[A}{}_{[G} T^{C]}{}_{H]BD}=
\frac{4}{3} \delta^{[A}{}_{[B} T^{C]}{}_{DGH]}
\end{equation}
So, because of equation (\ref{T-idbis}), we have:
\begin{equation}
\label{ABeq}
{2\over 3}\delta^{[A}{}_{[B} T^{C]}{}_{D]GH}
+{2\over 3}\delta^{[A}{}_{[G} T^{C]}{}_{H]BD}
=(u^{AB}{}_{\Omega\Sigma} + v^{AB\Omega\Sigma})
e_{\Omega\Sigma}^\alpha {\cal B}_{CDEF}
\end{equation}
By inserting the expressions of $T$ and $\cal B$ in
terms of the $t$'s into  equation (\ref{ABeq}) and collecting the coefficients
the independent scalar--field components, we retrieve equation (\ref{lt-id}).
\par
This concludes our proof. }
\end{enumerate}
}
\section{Solution of the algebraic $t$--identity}
The algebraic $t$--identity \eqn{lt-id} is a linear equation imposed
on the embedding matrix ${\cal E}$. We have solved it  by
means of a computer program and we have found the  36--parameter
solution displayed in tables \ref{matap1}, \ref{matap2},\ref{matap3},
\ref{matap4},\ref{matam1},\ref{matam2},\ref{matam3},\ref{matam4},
\ref{matacar}. In order to explain the content of these
tables we have to describe our notations and our construction in more
detail.
\par
\subsection{Embedding of the electric group}
The first information we need to specify is the explicit embedding of
the electric subalgebra $SL(8,\IR)$ into the U--duality algebra
$E_{7(7)}$. For this latter we adopt the conventions and notations of \cite{noi3}.
\subsubsection{The $E_{7(7)}$ algebra: roots and weights}
We consider the standard $E_7$ Dynkin diagram
(see fig. $3$) and we name $\alpha_i$ ($i=1,\dots,7$)
the corresponding simple roots.
\begin{figure}[ht]
\label{e7dyn}
\caption{$E_7$ Dynkin diagram and root labeling}
\centerline{\epsfig{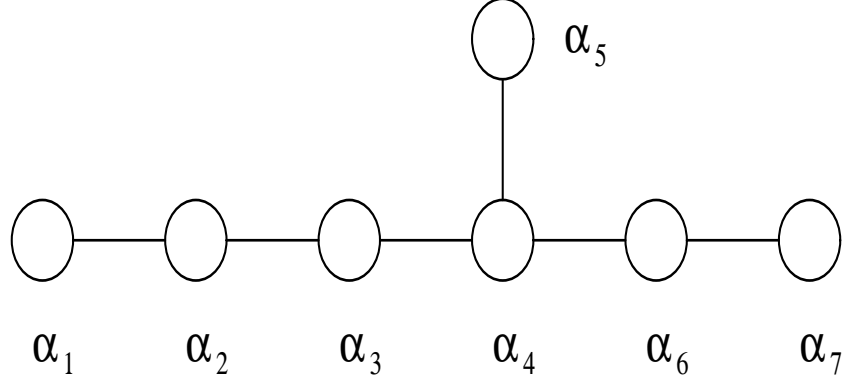}}
\end{figure}
\noindent
Having fixed this basis, each $E_{7(7)}$ root is intrinsically identified
by its Dynkin labels, namely by its integer valued components in the
simple root basis.
\par
Having identified the roots, the next step we need is the construction of
the real fundamental representation $SpD(56)$  of our
U--duality Lie algebra $E_{7(7)}$. For this we  need
the corresponding weight vectors ${\vec W}$.\par
A particularly relevant property of the maximally non--compact
real sections of a simple complex Lie algebra is that all
its irreducible representations are real.  $E_{7(7)}$ is the
maximally non compact real section of the complex Lie algebra $E_7$, hence
all its irreducible representations $\Gamma$  are real.
This implies that if an element of the  weight lattice ${\vec W} \, \in \, \Lambda_w$ is
a weight of a given irreducible representation
${\vec W}\in \Gamma$ then also its negative is a weight of the
same representation: $-{\vec W}\in \Gamma$. Indeed changing sign to
the weights corresponds to complex conjugation.
\par
According to standard Lie algebra lore
every irreducible representation of a simple Lie algebra $\IG$ is identified
by a unique {\it highest} weight ${\vec W}_{max}$.
Furthermore all weights can be expressed as
integral non--negative linear  combinations of the {\it simple}
weights ${\vec W}_{\ell}\,(\ell=1,...,r=\mbox{rank}(\IG)) $,
whose components are named the Dynkin labels of the weight.
The simple weights ${\vec W}_{i}$ of $\IG$ are the generators of the
dual lattice to the root lattice and are defined by the condition:
\begin{equation}
\frac{2 ({\vec W}_{i}\, ,\, {\vec \alpha}_{j})}{({\vec \alpha}_{j}\, ,\,
{\vec \alpha}_{j})}=\delta_{ij}
\end{equation}
In the simply laced $E_{7(7)}$ case, the previous equation
simplifies as follows
\begin{equation}
({\vec W}_{i}\, ,\, {\vec \alpha}_{j})=\delta_{ij}
\label{simw}
\end{equation}
where ${\vec \alpha}_{j}$ are the the simple roots.
Using the Dynkin diagram of
$E_{7(7)}$ (see fig. $3$) from eq.\eqn{simw} we can easily
obtain the explicit expression of the simple weights.
\par
The Dynkin labels of the highest weight of an irreducible
representation $\Gamma$ give  the Dynkin labels of the
representation. Therefore the representation is usually denoted by
$\Gamma[n_1,...,n_{r}]$. All the weights ${\vec W}$ belonging
to the representation $\Gamma$ can be described by $r$ integer
non--negative numbers $q^\ell$ defined by the following equation:
\begin{equation}
{\vec W}_{max}-{\vec W}=\sum_{\ell=1}^{r}q^\ell{\vec \alpha}_{\ell}
\label{qi}
\end{equation}
where $\alpha_\ell$ are the simple roots.
According to this standard formalism the fundamental real representation $SpD(56)$
of $E_{7(7)}$ is $\Gamma[1,0,0,0,0,0,0]$
and the expression of its weights in terms of $q^\ell$ is given in table
\ref{e7weight}, the highest weight being ${\vec W}^{(51)}$.
\par
We can now explain the specific ordering of the weights we have
adopted.
\par
First of all we have separated the $56$ weights in two
groups of $28$ elements so that the first group:
\begin{equation}
{\vec \Lambda}^{(n)}={\vec W}^{(n)} \quad n=1,...,28
\label{elecweight}
\end{equation}
are the weights for  the irreducible {\bf 28} dimensional representation
of the
{\sl electric} subgroup {\bf $SL(8,\IR) \subset E_{7(7)}$}.
The remaining group of $28$ weight vectors are the weights for the
transposed representation of the same group that we name ${\bf \bar{28}}$.
\par
Secondly the $28$ weights ${\vec \Lambda}$
have been arranged according to the decomposition with respect to the
T--{\it duality} subalgebra $SO(6,6)\subset E_7(7)$:  the first $16$
correspond to R--R vectors and are the weights of the spinor
representation of $SO(6,6)$ while the last $12$ are associated with N--S
fields and correspond to the weights of the vector representation of
$SO(6,6)$.
\subsubsection{The matrices of the fundamental {\bf 56}
representation}
Equipped with the weight vectors we can now proceed to the explicit
construction of the ${\bf SpD(56)}$ representation
of $E_{7(7)}$. In our construction the basis vectors
are the $56$ weights, according to the enumeration of table \ref{e7weight}.
What we need are the $56\times 56$ matrices associated with the  $7$
Cartan generators $H_{{\vec \alpha}_i}$ ($i=1,\dots , 7$) and with
the $126$ step operators $E^{\vec \alpha}$ that are defined by:
\begin{eqnarray}
\left[ SpD_{56}\left( H_{{\vec \alpha}_i} \right)\right]_{nm}&  \equiv &  \langle
{\vec W}^{(n)} \vert \,  H_{ {\vec \alpha}_i } \,\vert {\vec W}^{(m)}
\rangle \nonumber\\
\left[ SpD_{56}\left( E^{ {\vec \alpha} } \right)\right]_{nm}&  \equiv &  \langle
{\vec W}^{(n)} \vert \,  E^{ {\vec \alpha} }  \,\vert {\vec W}^{(m)}
\rangle
\label{sp56defmat}
\end{eqnarray}
Let us begin with the Cartan generators. As a basis of the
Cartan subalgebra we use the generators $H_{\vec \alpha_i}$ defined by the
commutators:
\begin{equation}
\left[ E^{{\vec \alpha }_i}, E^{-{\vec \alpha }_i} \right] \, \equiv
\, H_{\vec \alpha_i}
\label{cartbadefi}
\end{equation}
In the $SpD(56)$ representation the corresponding matrices are
diagonal and of the form:
\begin{equation}
\langle {\vec W}^{(p)} \vert\, H_{\vec \alpha_{i}}\, \vert
{\vec W}^{(q)}\rangle\, =\,\left({\vec W}^{(p)},{\vec \alpha_{i}}\right)
\delta_{p\, q}\quad ; \quad ( p,q\, =\, 1,...,56)
\label{cartane7}
\end{equation}
The scalar products
\begin{equation}
\left({\vec \Lambda}^{(n)} \, \cdot \, {\vec h},
-{\vec \Lambda}^{(m)} \, \cdot \, {\vec h}\right)\, =\, \left({\vec W}^{(p)}
 \, \cdot \,
{\vec h}\right) \quad ; \quad(n,m=1,...,28\, ;\, p=1,...,56)
\end{equation}
 are to be understood in the following way:
\begin{equation}
{\vec W}^{(p)} \, \cdot \, {\vec h}\, =\, \sum_{i=1}^{7}\left({\vec W}^{(p)},
{\vec \alpha_{i}}\right)h^i
\end{equation}
\par
Next we construct the matrices associated with the step operators. Here the first
observation is that it suffices to consider the positive roots. Because
of the reality of the representation, the matrix associated with the
negative of a root is just the transposed of that associated with the
root itself:
\begin{equation}
E^{-\alpha} = \left[ E^\alpha \right]^T \, \leftrightarrow \, \langle
{\vec W}^{(n)} \vert \,  E^{- {\vec \alpha} }  \,\vert {\vec W}^{(m)} \rangle \, = \,
\langle {\vec W}^{(m)} \vert \,  E^{ {\vec \alpha} }  \,\vert {\vec W}^{(n)} \rangle
\label{transpopro}
\end{equation}
The method we have followed to obtain the matrices for all the
positive roots is that of constructing first the $56\times 56$
matrices for the step operators $E^{\vec \alpha_{\ell}}\, (\ell=1,...,7)$
associated with the simple roots and then generating all the others
through their commutators. The construction rules for the $SpD(56)$
representation of  the six operators $E^{\alpha_{\ell}} \, (\ell\neq 5)$
are:
\begin{equation}
\ell \, \neq \, 5 \quad \,
\Biggl \{ \matrix {\langle {\vec W}^{(n)} \vert\, E^{\vec \alpha_{\ell}}\, \vert
{\vec W}^{(m)}\rangle & = & \delta_{{\vec W}^{(n)},
{\vec W}^{(m)}+{\vec \alpha}_\ell} &;& n,m=1,\dots, 28 \cr
\langle {\vec W}^{(n+28)} \vert\, E^{\vec \alpha_{\ell}}\, \vert
{\vec W}^{(m+28)}\rangle & = & -\delta_{{\vec W}^{(n+28)},
{\vec W}^{(m+28)}+{\vec \alpha}_\ell} & ; & n,m=1,\dots, 28 \cr }
\label{repineq5}
\end{equation}
The six simple roots ${\vec \alpha_{\ell}} $ with $ \ell \neq 5$
belong also to the the Dynkin diagram of the electric
subgroup {\bf SL(8,\IR)}.  Thus their shift
operators have a block diagonal action on the {\bf 28} and ${\bf \bar{28}}$
subspaces of the $SpD(56)$ representation that are irreducible
under the electric subgroup. From eq.\eqn{repineq5} we conclude
that:
\begin{equation}
\ell \, \neq \, 5 \quad \,SpD_{56}\left(E^{{\vec \alpha}_\ell} \right)=
\left(\matrix { A[{{\vec \alpha}_\ell}]
& {\bf 0} \cr {\bf 0} & - A^T[{{\vec \alpha}_\ell}] \cr} \right)
\label{spdno5}
\end{equation}
the $28 \times 28$ block  $A[{{\vec \alpha}_\ell}]$ being defined
by the first line of eq.\eqn{repineq5}.\par
On the contrary the operator $E^{\vec \alpha_{5}}$, corresponding to the only
root of the $E_7$ Dynkin diagram that is not also part of the $A_7$
diagram is represented by a matrix whose non--vanishing $28\times 28$ blocks
are off--diagonal. We have
\begin{equation}
SpD_{56}\left(E^{{\vec \alpha}_5} \right)=\left(\matrix { {\bf 0} & B[{{\vec \alpha}_5}]
\cr  C[{{\vec \alpha}_5}] & {\bf 0} \cr} \right)
\label{spdyes5}
\end{equation}
where both $B[{{\vec \alpha}_5}]=B^T[{{\vec \alpha}_5}]$ and
$C[{{\vec \alpha}_5}]=C^T[{{\vec \alpha}_5}]$ are symmetric $28
\times 28$ matrices. More explicitly the  matrix
 $SpD_{56}\left(E^{{\vec \alpha}_5} \right)$
is given by:
\begin{eqnarray}
&& \langle {\vec W}^{(n)} \vert\, E^{\vec \alpha_{5}}\, \vert
{\vec W}^{(m+28)}\rangle \, =\,  \langle {\vec W}^{(m)} \vert\,
E^{\vec \alpha_{5}}\, \vert {\vec W}^{(n+28)}\rangle \nonumber \\
&& \langle {\vec W}^{(n+28)} \vert\, E^{\vec \alpha_{5}}\, \vert
{\vec W}^{(m)}\rangle \, =\,  \langle {\vec W}^{(m+28)} \vert\,
E^{\vec \alpha_{5}}\, \vert {\vec W}^{(n)}\rangle
\label{sim5}
\end{eqnarray}
with
\begin{equation}
\begin{array}{rcrcrcrcrcr}
  \langle {\vec W}^{(7)} \vert\, E^{\vec \alpha_{5}}\, \vert
{\vec W}^{(44)}\rangle & =& -1 & \null &
  \langle {\vec W}^{(8)} \vert\, E^{\vec \alpha_{5}}\, \vert
{\vec W}^{(42)}\rangle & = & 1  & \null &
   \langle {\vec W}^{(9)} \vert\, E^{\vec \alpha_{5}}\, \vert
{\vec W}^{(43)}\rangle & = & -1  \\
   \langle {\vec W}^{(14)} \vert\, E^{\vec \alpha_{5}}\, \vert
{\vec W}^{(36)}\rangle & = & 1 & \null &
   \langle {\vec W}^{(15)} \vert\, E^{\vec \alpha_{5}}\, \vert
{\vec W}^{(37)}\rangle & = & -1 & \null &
 \langle {\vec W}^{(16)} \vert\, E^{\vec \alpha_{5}}\, \vert
{\vec W}^{(35)}\rangle & = & -1  \\
   \langle {\vec W}^{(29)} \vert\, E^{\vec \alpha_{5}}\, \vert
{\vec W}^{(6)}\rangle  & = & -1 & \null &
   \langle {\vec W}^{(34)} \vert\, E^{\vec \alpha_{5}}\, \vert
{\vec W}^{(1)}\rangle  & = & -1 & \null &
  \langle {\vec W}^{(49)} \vert\, E^{\vec \alpha_{5}}\, \vert
{\vec W}^{(28)}\rangle & = & 1 \\
  \langle {\vec W}^{(50)} \vert\, E^{\vec \alpha_{5}}\, \vert
{\vec W}^{(27)}\rangle & = & -1 & \null &
   \langle {\vec W}^{(55)} \vert\, E^{\vec \alpha_{5}}\, \vert
{\vec W}^{(22)}\rangle & = &  -1 & \null &
   \langle {\vec W}^{(56)} \vert\, E^{\vec \alpha_{5}}\, \vert
{\vec W}^{(21)}\rangle & =& 1 \\
\end{array}
\label{sim5bis}
\end{equation}
In this way we have completed the construction of the $E^{{\vec \alpha}_\ell}$
operators associated with simple roots. For the matrices associated
with higher roots we just proceed iteratively in the following way.
As usual we organize the roots by height :
\begin{equation}
{\vec \alpha}=n^\ell \, {\vec \alpha}_\ell \quad \rightarrow
\quad \mbox{ht}\,{\vec \alpha} \, = \, \sum_{\ell=1}^{7} n^\ell
\label{altezza}
\end{equation}
and for the roots $\alpha_i + \alpha_j$ of height $\mbox{ht}=2$ we
set:
\begin{equation}
SpD_{56} \left( E^{ \alpha _i + \alpha _j} \right) \equiv \left[
SpD_{56}\left(E^{\alpha _i} \right) \, , \,
SpD_{56}\left(E^{\alpha _i} \right) \right] \quad ; \quad i<j
\label{alto2}
\end{equation}
Next for the roots of $\mbox{ht}=3$ that can be written as $\alpha_i
+ \beta $ where $\alpha_i$ is simple and $\mbox{ht}\, \beta\, =\, 2$
we write:
\begin{equation}
SpD_{56} \left( E^{ \alpha _i + \beta} \right) \equiv \left[
SpD_{56}\left(E^{\alpha _i} \right) \, , \,
SpD_{56}\left(E^{\beta} \right) \right]
\label{alto3}
\end{equation}
Obtained the matrices for the roots of $\mbox{ht}=3$ one proceeds in
a similar way for those of the next height and so on up to exhaustion
of all the $63$ positive roots.
\par
This concludes our description of the algorithm by means of which our
computer program constructed all the $133$ matrices spanning the
$E_{7(7)}$ Lie algebra  in the $SpD(56)$ representation.  A fortiori,
if we specify the embedding we have the matrices generating the
electric subgroup $SL(8,\IR)$.
\subsubsection{The $SL(8,\IR)$ subalgebra}
\par
The Electric $Sl(8,\IR)$ subalgebra is identified in $E_{7(7)}$ by
specifying its simple roots $\beta_i$ spanning the standard $A_7$
Dynkin diagram. The Cartan generators are the same for the
$E_{7(7)}$ Lie algebra as for the $SL(8,\IR)$ subalgebra and if we
give $\beta_i$ every other generator is defined.
The basis we have chosen is the following one:
\begin{equation}
\begin{array}{ccccccc}
\beta _1\,&=&\,\alpha _2+2\alpha _3+3\alpha _4+2\alpha _5+2 \alpha _6+
\alpha _7
&;&
\beta _2 &=& \alpha_1 \cr
\beta _3\,&=&\,\alpha _2 & ; & \beta _4 & =&  \alpha _3 \cr
\beta _5\,&=&\,\alpha _4 &; & \beta _6 & =& \alpha _6 \cr
\beta _7\,&=&\,\alpha_7  & \null & \null & \null & \null \cr
\end{array}
\label{simrot}
\end{equation}
The complete set of positive roots of $SL(8\IR)$ is then composed
of $28$ elements that we name $\rho_i$ ($i=1,\dots,28$) and that are
enumerated according to our chosen order in table \ref{h_igauging}.
\begin{table}[ht]\caption{{\bf}
The choice of the order of the $SL(8,\IR)$ roots:}
\label{h_igauging}
\begin{center}
\begin{tabular}{||cl||}
\hline
\hline
\null & \null \\
$ \rho_1 \, \equiv\, $ &$ \beta_2  $ \\
$ \rho_2 \, \equiv\, $ &$ \beta_2+\beta_3  $   \\
$ \rho_3 \, \equiv\, $ &$ \beta_2+\beta_3+\beta_4 $ \\
$ \rho_4 \, \equiv\, $ &$ \beta_2+\beta_3+\beta_4+\beta_5  $    \\
$ \rho_5 \, \equiv\, $ &$ \beta_2+\beta_3+\beta_4+\beta_5+\beta_6  $ \\
$ \rho_6 \, \equiv\, $ &$ \beta_3 $    \\
$ \rho_7 \, \equiv\, $ &$ \beta_3+\beta_4 $  \\
$ \rho_8 \, \equiv\, $ &$ \beta_3+\beta_4+\beta_5 $    \\
$ \rho_9 \, \equiv\, $ &$ \beta_3+\beta_4+\beta_5+\beta_6  $  \\
$ \rho_{10} \, \equiv\, $ &$ \beta_4 $   \\
$ \rho_{11} \, \equiv\, $ &$ \beta_4+\beta_5 $  \\
$ \rho_{12} \, \equiv\, $ &$ \beta_4+\beta_5+\beta_6   $    \\
$ \rho_{13} \, \equiv\, $ &$ \beta_5 $ \\
$ \rho_{14} \, \equiv\, $ &$ \beta_5+\beta_6  $    \\
$ \rho_{15} \, \equiv\, $ &$ \beta_6   $  \\
$ \rho_{16} \, \equiv\, $ &$ \beta_1+\beta_2+\beta_3+\beta_4+\beta_5+\beta_6+\beta_7  $    \\
$ \rho_{17} \, \equiv\, $ &$ \beta_2+\beta_3+\beta_4+\beta_5+\beta_6+\beta_7  $  \\
$ \rho_{18} \, \equiv\, $ &$ \beta_3+\beta_4+\beta_5+\beta_6+\beta_7   $    \\
$ \rho_{19} \, \equiv\, $ &$ \beta_4+\beta_5+\beta_6+\beta_7  $ \\
$ \rho_{20} \, \equiv\, $ &$ \beta_5+\beta_6+\beta_7  $    \\
$ \rho_{21} \, \equiv\, $ &$ \beta_6+\beta_7  $\\
$ \rho_{22} \, \equiv\, $ &$ \beta_1  $    \\
$ \rho_{23} \, \equiv\, $ &$ \beta_1+\beta_2  $ \\
$ \rho_{24} \, \equiv\, $ &$ \beta_1+\beta_2+\beta_3 $    \\
$ \rho_{25} \, \equiv\, $ &$ \beta_1+\beta_2+\beta_3+\beta_4  $  \\
$ \rho_{26} \, \equiv\, $ &$ \beta_1+\beta_2+\beta_3+\beta_4+\beta_5   $    \\
$ \rho_{27} \, \equiv\, $ &$ \beta_1+\beta_2+\beta_3+\beta_4+\beta_5+\beta_6  $ \\
$ \rho_{28} \, \equiv\, $ &$ \beta_7 $ \\
\null & \null         \\
\hline
\end{tabular}
\end{center}
\end{table}
\par
Hence the $63$ generators of the $SL(8,\IR)$ Lie algebra are:
\begin{equation}
\begin{array}{ccc}
\mbox{The 7 Cartan generators} & C_i   =   H_{\alpha_i} & i=1,\dots,7\cr
\mbox{The 28 positive root generators} & E^{\rho_i} & i=1,\dots,28
\cr
\mbox{The 28 negative root generators} & E^{-\rho_i} & i=1,\dots,28
\cr
\end{array}
\label{genenume}
\end{equation}
and since the $56 \times 56$ matrix representation of each $E_{7(7)}$
Cartan generator or step operator was constructed in the previous
subsection it is obvious that it is in particular given for the subset
of those that belong to the $SL(8,\IR)$ subalgebra. The basis of this
matrix representation is provided by the weights enumerated in table
\ref{e7weight}. \par
In this way we have concluded our illustration of the basis in which
we have solved the algebraic $t$--identity.
The result is the $28 \times 63 $ matrix:
\begin{equation}
{\cal E}(h,\ell) \, \longrightarrow \, e^{\alpha}_{W}(h,\ell)
 \label{solvomat}
\end{equation}
where the index $W$ runs on the $28$ negative weights of table
\ref{e7weight}, while the index $\alpha$ runs on all the $SL(8,\IR)$
generators according to eq. \eqn{genenume}. The matrix ${\cal E}(h,p)$
depends on $36$ parameters that we have named:
\begin{equation}
\begin{array}{cc}
h_i & i=1,\dots, 8 \cr
\ell_i & i=1, \dots ,28 \cr
\end{array}
\label{hpparam}
\end{equation}
and its entries are explicitly displayed in tables
\ref{matap1}, \ref{matap2},\ref{matap3},
\ref{matap4},\ref{matam1},\ref{matam2},\ref{matam3},\ref{matam4},
\ref{matacar} as we have already stressed at the beginning of this
section. The distinction between the $h_i$ parameters  and
the $\ell_i$ parameters  has been drawn in the following way:
\begin{itemize}
\item The $8$ parameters  $h_i$ are those that never multiply a Cartan
generator
\item The $28$ parameters $\ell_i$ are those that multiply at least one
Cartan generator.
\end{itemize}
In other words if we set all the $\ell_i=0$ the gauge subalgebra
$G_{gauge} \subset SL(8,\IR)$ is composed solely of step operators
while if you switch on also the $\ell_i$.s then some Cartan generators
appear in the Lie algebra. This distinction will be very useful
in classifying the independent solutions.
\section{Classification of gauged $N=8$ supergravities}
Equipped with the explicit solution of the $t$--identity encoded
in the embedding matrix ${\cal E}$ we can now address the
complete classification of the gauged supergravity models.
As already anticipated in the introduction our result is that
the complete set of possible theories coincides with the gaugings
found by Hull \cite{hull} (together with the ones simply outlined by 
Hull \cite{hull})
in the middle of the eighties and correspond to all possible
non--compact real forms of the $SO(8)$ Lie algebra plus a number
of Inonu--Wigner contractions thereof. Since Hull's
method to obtain these gaugings was not based on an exhaustive
analysis the doubt existed whether his set of theories was complete
or not. Our analysis shows that in fact it was. Furthermore since
our method is constructive it provides the means to study in a
systematic way the features of all these models. In this
section we derive such a result.
\par
We have to begin our discussion with two observations:
\begin{enumerate}
\item{ The solution of $t$--identities encoded in the matrix ${\cal
E}(h,\ell)$ is certainly overcomplete since we are still free to
conjugate any gauge algebra $G_{gauge}$ with an arbitrary finite
element of the electric group $g \, \in \, SL(8,\IR)$:
$G^\prime_{gauge}=g \,G_{gauge} \, g^{-1} $ yields a completely
physically equivalent gauging as $G_{gauge}$. This means that we
need to consider the $SL(8,\IR)$ transformations of the matrix
${\cal E}(h,\ell)$ defined as:
\begin{equation}
\forall \, g \,\in \, SL(8,\IR) \, : \,  g  \cdot {\cal E}(h,\ell)
\equiv D_{28}(g^{-1}) \, {\cal E}(h,\ell) \, D_{63}(g)
\label{conjug}
\end{equation}
where $D_{28}(g)$ and $D_{63}(g)$ denote the matrices of the ${\bf 28}$
and the ${\bf 63}$ representation respectively.
If two set of parameters $\{h,\ell\}$ and
$\{ h^\prime,\ell^\prime \}$ are related by an $SL(8,\IR)$
conjugation, in the sense that:
\begin{equation}
 \exists g \, \in \,SL(8,\IR) \, : \,
 {\cal E}(h^\prime,\ell^\prime)=g \cdot {\cal E}(h,\ell)
 \label{modding}
\end{equation}
then the theories described by  $\{h,\ell\}$  and
$\{ h^\prime,\ell^\prime \}$ are the same theory. In other words what
we need is the space of orbits of $SL(8,\IR)$ inequivalent embedding
matrices.}
\item {Possible theories obtained by choosing a set of $\{h,\ell\}$
parameters are further restricted by the constraints that
\begin{itemize}
\item The selected generators of $SL(8,\IR)$ should close a Lie
subalgebra $G_{gauge}$
\item The selected vectors (=weights, see table \ref{e7weight}) should 
transform in the
coadjoint representation $Coadj \left(G_{gauge}\right)$
\end{itemize}
}
\end{enumerate}
In view of these observations a natural question we should pose is the following one:
{\sl is there a natural way to understand why the number of
parameters on which the embedding matrix depends is, a part from an immaterial overall
constant, precisely 35?}.
The answer is immediate and inspiring. Because of point 2) in the
above list of properties the $28$ linear combinations of $SL(8,\IR)$
generators:
\begin{equation}
  T_W \equiv e_{W}^{\phantom{W}\alpha} \left(h,\ell\right) \, G_\alpha
\label{wgene}
\end{equation}
must span the adjoint representation of a $28$--dimensional 
subalgebra $G_{gauge}(h,\ell)$ of
the \break \hfill $SL(8,\IR)$ algebra. \par \noindent Naming ${\cal G}_{gauge}(h,\ell) $
the corresponding Lie subgroup, because of its very definition we
have that the matrix ${\cal E}(h,\ell)$ is invariant under
transformations of ${\cal G}_{gauge}(h,\ell)$
\footnote{Note that some of the 28 generators of ${\cal G}_{gauge}(h,\ell)  \, \subset \, SL(8,\IR)$
may be  represented trivially in the adjoint representation, but in this case also the
corresponding group transformations leave  the embedding matrix invariant.}:
\begin{flushleft}
\begin{equation}
\forall \, \gamma \, \in  \, {\cal G}_{gauge}(h,\ell\,  \subset \, SL(8,\IR): \quad
\gamma \, \cdot \, {\cal E}(h,\ell) = {\cal E}(h,\ell)
\label{invarialetto}
\end{equation}
\end{flushleft}
Comparing eq.\eqn{invarialetto} with eq.\eqn{conjug} we see that
having fixed a matrix  ${\cal E}(h,\ell)$ and hence an algebra ${\cal G}_{gauge}(h,\ell)$,
according to point 1) of the above discussion
we can obtain a $35$--dimensional orbit of equivalent embedding matrices:
\begin{equation}
 {\cal E}\left( h^\prime(\mu),\ell^\prime(\mu)\right )\,  \equiv\, g(\mu)\,
 \cdot \, {\cal E}(h,\ell)  \quad \mbox{where} \quad g(\mu) \, \in \,
 \frac{SL(8,\IR)}{{\cal G}_{gauge}(h,\ell)}
 \label{quozie}
\end{equation}
Hence, $35 = 63 -28$ is the dimension of the coset manifold ${SL(8,\IR)}/{{\cal G}_{gauge}(h,\ell)}$
and \break \hfill ${\cal E}\left( h^\prime(\mu),\ell^\prime(\mu)\right )$ is
the embedding matrix for the family of conjugated, isomorphic, Lie
algebras:
\begin{equation}
  {\cal G}_{gauge}\left( h^\prime(\mu),\ell^\prime(\mu)\right )\, =
  g^{-1}(\mu) \,{{\cal G}_{gauge}(h,\ell)} \, g(\mu)
  \label{buscarmatri}
\end{equation}
An essential and a priori unexpected conclusion follows from this discussion.
\par
\bpropo
{\large The gauged $N=8$ supergravity models cannot depend on more than a
single continuous parameter  (=coupling constant), even if they
correspond to gauging a multidimensional abelian algebra}
\epropo
\bpr{-}
\epr
{\small Since the explicit solution of the algebraic $t$--identities has produced
an embedding matrix ${\cal E}(h,\ell)$ depending on no more than $36$--parameters, then the only
continuous parameter which is physically relevant is the overall
proportionality constant. The remaining $35$--parameters can be
reabsorbed by $SL(8,\IR)$ conjugations according to eq.\eqn{buscarmatri}}
\vskip 0.2cm
\par
In other words what we have found is that the space of orbits we are looking for is a discrete
space. The classifications of gauged supergravity models is just a
classification of gauge algebras a single coupling constant being
assigned to each case. This is considerably different from other supergravities
with less supersymmetries, like the $N=2$ case. There gauging a group $G_{gauge}$ involves
as many coupling constants as there are simple factors in
$G_{gauge}$. So in those cases not only we have a much wider variety
of possible gauge algebras but also we have lagrangians depending on
several continuous parameters. In the $N=8$ case supersymmetry constraints
the theory in a much stronger way. We want to stress that this
is an yield of supersymmetry and not of Lie algebra theory. It is the algebraic $t$--identity, enforced by the closure of Bianchi
identities, that admits a general solution depending only on $36$--parameters.
If the solution depended on $35 + m$   parameters then we might have
introduced $m$ relevant continuous parameters into the Lagrangian.
\par
Relying on these observations we are left with the problem of
classifying the orbit space already knowing that it is composed of
finite number of discrete elements. Orbits are characterized in terms
of invariants, so we have to ask ourselves what is the natural
invariant associated with the embedding matrix ${\cal E}(h,\ell)$.
The answer is once again very simple. It is the {\it signature} of the
{\it Killing--Cartan 2--form} for the resulting gauge algebra ${\cal G}_{gauge}(h,\ell)$.
Consider the $28$ generators \eqn{wgene} and define:
\begin{equation}
\begin{array}{rcl}
\eta_{W_1W_2}\left( h,\ell\right ) & \equiv & \mbox{Tr}  \, \left( T_{W_1}
\, T_{W_1} \right)   \cr
\null & = & e_{W_1}^{\phantom{W_1}\alpha }\left(h,\ell \right ) \, e_{W_2}^{\phantom{W_1}\beta }\left(h,\ell\right) \,
\mbox{Tr}  \,
\left( G_ \alpha \, G_ \beta \right) \cr
\null & =  & e_{W_1}^{\phantom{W_1}\alpha }\left(h,\ell\right) \, e_{W_2}^{\phantom{W_1}\beta }\left(h,\ell\right)
\, B_{\alpha\beta}\cr
\end{array}
\label{assas1}
\end{equation}
where the trace $Tr$ is taken over any representation and the
constant matrix $ B_{\alpha\beta} \equiv \mbox{Tr} \,\left( G_ \alpha \, G_ \beta \right)$ is the Killing--Cartan
2--form of the $SL(2,\IR)$ Lie algebra. The $ 28 \times 28 $ matrix
is the Killing--Cartan 2--form of the gauge algebra $G_{gauge}$. As
it is well known from general Lie algebra theory, by means of
suitable changes of bases inside the same Lie algebra the matrix $\eta_{W_1W_2}\left( h,\ell\right )$
can be diagonalized and its eigenvalues can be reduced to be either of
modulus one or zero. What cannot be done since it corresponds to an
intrinsic characterization of the Lie algebra is to change the
signature of $\eta_{W_1W_2}\left( h,\ell\right )$, namely the ordered set
of $28$ signs (or zeros) appearing on the principal diagonal when $\eta_{W_1W_2}\left( h,\ell\right )$
is reduced to diagonal form. Hence what is constant throughout   an
$SL(8,\IR)$ orbit is the signature. Let us   name $\Sigma \left( \mbox{orbit} \right)$ the
$28$ dimensional vector characterizing the signature of an orbit:
\begin{equation}
\Sigma\left( \mbox{orbit} \right ) \,\equiv\, \mbox{signature} \, \left[
\eta_{W_1W_2}\left( h^\prime(\mu),\ell^\prime(\mu)\right )\, \right]
\label{segnato}
\end{equation}
>From our discussion we conclude that
\bpropo
{\large
The classification of gauged N=8 models has been reduced to the
classification of the signature vectors $\Sigma\left( \mbox{orbit} \right )$ of
eq.\eqn{segnato}}
\epropo
The procedure to calculate $\Sigma\left( \mbox{orbit} \right )$
associated with an orbit $\eta_{W_1W_2}\left( h^\prime(\mu),\ell^\prime(\mu)\right )\, $ is
that of choosing the representative $\left( h^\prime(\mu_0),\ell^\prime(\mu_0)\right )$ for
which the corresponding matrix   $\eta_{W_1W_2}\bigl ( h^\prime(\mu_0),$
$\ell^\prime(\mu_0)\bigr )\,$
is diagonal and then to evaluate the signs of the diagonal elements.
In principle finding the appropriate $h^\prime(\mu_0),\ell^\prime(\mu_0)$ could be a difficult
task since we are supposed to diagonalize a $28 \times 28$ matrix.
However our choice of coordinates on the parameter space is such that
  our task becomes very simple. Using the results of
tables \ref{matap1}, \ref{matap2},\ref{matap3},
\ref{matap4},\ref{matam1},\ref{matam2},\ref{matam3},\ref{matam4},
\ref{matacar}   we can calculate the matrix
$\eta_{W_1W_2}\left( h,\ell \right )\,$ and for generic   values of $h_i$
and $\ell_i$ we find that all of its $28 \times 28 $ entries are non vanishing; yet
setting $\ell_i=0$ the matrix becomes automatically diagonal and we get:
\begin{equation}
\begin{array}{rlcccccccl}
\eta\left( h,\ell = 0\right ) \,= & \mbox{diag}
\Bigl \{
 & - {h_7}\,{h_8}  , &{h_1}\,{h_6}, &{h_2}\,{h_6},&
   - {h_3}\,{h_6} & ,{h_4}\,{h_6},& - {h_5}\,{h_6}  ,  &
   {h_1}\,{h_2}, & \null \cr
     \null & \null & - {h_1}\,{h_3}  , & {h_1}\,{h_4},   &
   - {h_1}\,{h_5}  , & - {h_2}\,{h_5}  , & {h_3}\,{h_5},
 &   - {h_4}\,{h_5}  , & {h_2}\,{h_4}, & \null \cr
   \null & \null & - {h_2}\,{h_3}  ,  &
   - {h_3}\,{h_4}  , & {h_1}\,{h_7}, & {h_2}\,{h_7},    &
   - {h_3}\,{h_7}  ,& {h_4}\,{h_7}, & - {h_5}\,{h_7}, & \null  \cr
   \null & \null &   {h_6}\,{h_7}, & - {h_1}\,{h_8}  ,& - {h_2}\,{h_8}  ,  &
   {h_3}\,{h_8}, &- {h_4}\,{h_8}  , & {h_5}\,{h_8},
 &  - {h_6}\,{h_8} & \Bigr \} \cr
 \end{array}
\end{equation}
Hence all possible signatures $\Sigma \left(orbit \right)$ are
obtained by assigning to the parameters $h_i$ the values $1,-1,0$ in all possible ways.
Given an $h$ vector constructed in this way we have then to check that the corresponding $28$
generators \eqn{wgene} close a Lie subalgebra and accept only those
for which this happens. Clearly such an algorithm can be easily implemented by means of
a computer program. The result is provided by a table of $SL(8,\IR)$ Lie subalgebras
identified by a corresponding acceptable $h$--vector. This result is
displayed in the table that follows (see eq.\eqn{risulato}). In this
table in addition to the $h$--vector that identifies it we have
displayed the signature of the Killing--Cartan form by writing the
numbers $n_+$,$n_-$,$n_0$ of its positive, negative and zero
eigenvalues. In addition we have also written the actual dimension of
the gauge algebra namely the number of generators that have a
non--vanishing representations or correspondingly the number of
gauged vectors that are gauged (=paired to a non vanishing
generator):
\begin{equation}
\begin{array}{|c|c|c|c|c|c|}
\hline
\mbox{Algebra} & {n_+} & {n_-} & {n_0}& \{h_1,h_2,h_3,h_4,h_5,h_6,h_7\} &
\mbox{dimension}
\cr
\hline
\hline
  SO( 8 ) & 28 & 0 & 0 & \{1,1,-1,1,-1,1,1,-1\} & 28 \cr
  SO( 1,7 ) & 21 & 7 & 0 & \{1,1,-1,1,-1,1,1,1\}& 28 \cr
  SO( 2,6 ) & 16 & 12 & 0 & \{-1,1,-1,1,-1,1,1,1\}& 28\cr
  SO( 3,5 ) & 13 & 15 & 0 & \{-1,-1,-1,1,-1,1,1,1\}& 28 \cr
  SO( 4,4 ) & 12 & 16 & 0 & \{-1,-1,1,1,-1,1,1,1\} & 28\cr
  SO(5,3  ) & 13 & 15 & 0 & \{-1,-1,1,-1,-1,1,1,1\}& 28\cr
  SO( 6,2 ) & 16 & 12 & 0 & \{-1,-1,1,-1,1,1,1,1\} & 28\cr
  SO( 7,1 ) & 21 & 7 & 0 & \{-1,-1,1,-1,1,-1,1,1\} & 28\cr
  \hline
  CSO( 1 , 7) & 0 & 0 & 28 & \{0,0,0,0,0,0,0,1\}& 7 \cr
  CSO(  2 , 6) & 1 & 0 & 27 & \{-1,0,0,0,0,0,0,1\} & 13 \cr
  CSO(  3 ,5 ) & 3 & 0 & 25 & \{-1,-1,0,0,0,0,0,1\}& 18 \cr
  CSO(  4 ,4 ) & 6 & 0 & 22 & \{-1,-1,1,0,0,0,0,1\}& 22 \cr
  CSO( 5  , 3) & 10 & 0 & 18 & \{-1,-1,1,-1,0,0,0,1\}& 25 \cr
  CSO( 6  , 2) & 15 & 0 & 13 & \{-1,-1,1,-1,1,0,0,1\}& 27\cr
  CSO(  7 , 1) & 21 & 0 & 7 & \{-1,-1,1,-1,1,-1,0,1\}& 28\cr
  \hline
  CSO(  1 ,1 , 6) & 0 & 1 & 27 &\{1,0,0,0,0,0,0,1\}& 13 \cr
  CSO(  1 ,2 , 5) & 1 & 2 & 25 & \{1,-1,0,0,0,0,0,1\} & 18\cr
  CSO(  2 ,1 , 5) & 1 & 2 & 25 & \{1,1,0,0,0,0,0,1\} & 18\cr
  CSO( 1  ,3, 4) & 3 & 3 & 22 & \{1,-1,1,0,0,0,0,1\}& 22\cr
  CSO( 2  ,2 , 4) & 2 & 4 & 22 & \{1,1,1,0,0,0,0,1\}  &  22  \cr
  CSO(  3 ,1, 4) & 3 & 3 & 22 & \{1,1,-1,0,0,0,0,1\} & 22\cr
  CSO(  1 ,4,3 ) & 6 & 4 & 18 & \{1,-1,1,-1,0,0,0,1\} & 25\cr
  CSO(  2 ,3 , 3) & 4 & 6 & 18 &\{1,1,1,-1,0,0,0,1\} & 25\cr
  CSO(  3 ,2, 3) & 4 & 6 & 18 & \{1,1,-1,-1,0,0,0,1\} & 25 \cr
  CSO(  4 ,1, 3) & 6 & 4 & 18 & \{1,1,-1,1,0,0,0,1\} & 25\cr
  CSO(  1 ,5 ,2 ) & 10 & 5 & 13 & \{1,-1,1,-1,1,0,0,1\} & 27\cr
  CSO(  2 ,4, 2) & 7 & 8 & 13 & \{1,1,1,-1,1,0,0,1\} & 27 \cr
  CSO( 3  ,3,2 ) & 6 & 9 & 13 & \{1,1,-1,-1,1,0,0,1\} & 27 \cr
  CSO(  4 ,2,2 ) & 7 & 8 & 13 & \{1,1,-1,1,1,0,0,1\} & 27\cr
  CSO(  5 ,1, 2) & 10 & 5 & 13 & \{1,1,-1,1,-1,0,0,1\} & 27\cr
  CSO(  1 ,6, 1) & 15 & 6 & 7 &\{1,-1,1,-1,1,-1,0,1\} & 28 \cr
  CSO(  2 ,5 ,1 ) & 11 & 10 & 7 &\{1,1,1,-1,1,-1,0,1\} & 28 \cr
l
  CSO( 3  ,4, 1) & 9 & 12 & 7 & \{1,1,-1,-1,1,-1,0,1\}& 28 \cr
  CSO( 4  ,3,1 ) & 9 & 12 & 7 & \{1,1,-1,1,1,-1,0,1\} & 28 \cr
  CSO(  5 ,2, 1) & 11 & 10 & 7 & \{1,1,-1,1,-1,-1,0,1\} & 28 \cr
  CSO( 6  ,1, 1) & 15 & 6 &  7   &\{1,1,-1,1,-1,1,0,1\}& 28 \cr
\hline
\hline
\end{array}
\label{risulato}
\end{equation}
By restricting the matrix  $e_{W}^{\phantom{W} \alpha}$ to the parameters
$h_i$ we can immediately write the correspondence between the vectors ${\vec W}^{(28+i)}$
 and the generators of the gauge algebra that applies to all the gaugings we have classified above.
 For the reader's convenience this correspondence is summarized in the 
 following table,
 where it suffices to substitute the corresponding values of $h_i$ to
 obtain the generators of each gauge algebra expressed as linear
 combinations of the $56$ positive and negative root step operators
 of $SL(8,\IR)$.
\begin{center}
\begin{tabular}{||cl||}
\hline
\hline
    Electric & \qquad \qquad\qquad Gauge \\
      vector &  \qquad\qquad\qquad generator\\
\hline
\null & \null \\
$ {\vec W}^{(35)} \, \leftrightarrow\, $ &$ h_2 E_{-\beta_2}-h_1
E_{\beta_2} $ \\
$ {\vec W}^{(36)} \, \leftrightarrow\, $ &$ h_3 E_{-\beta_2-\beta_3}+h_1 E_{\beta_2+\beta_3}  $   \\
$ {\vec W}^{(37)} \, \leftrightarrow\, $ &$ h_4 E_{-\beta_2-\beta_3-\beta_4}-h_1 E_{\beta_2+\beta_3+\beta_4} $ \\
$ {\vec W}^{(38)} \, \leftrightarrow\, $ &$ h_5 E_{-\beta_2-\beta_3-\beta_4-\beta_5}+h_1 E_{\beta_2+\beta_3+\beta_4+
\beta_5}  $    \\
$ {\vec W}^{(30)} \, \leftrightarrow\, $ &$ h_6 E_{-\beta_2-\beta_3-\beta_4-\beta_5-\beta_6}-h_1 E_{\beta_2+\beta_3+
\beta_4+\beta_5+\beta_6}  $ \\
$ {\vec W}^{(45)} \, \leftrightarrow\, $ &$ -h_7 E_{-\beta_2-\beta_3-\beta_4-\beta_5-\beta_6-\beta_7}+h_1 E_{\beta_2+
\beta_3+\beta_4+\beta_5+\beta_6+\beta_7} $    \\
$ {\vec W}^{(51)} \, \leftrightarrow\, $ &$h_1 E_{-\beta_1}+h_8
E_{\beta_1}  $  \\
$ {\vec W}^{(52)} \, \leftrightarrow\, $ &$ h_2 E_{-\beta_1-\beta_2}+h_8 E_{\beta_1+\beta_2} $    \\
$ {\vec W}^{(53)} \, \leftrightarrow\, $ &$  h_3 E_{-\beta_1-\beta_2-\beta_3}-h_8 E_{\beta_1+\beta_2+\beta_3}  $  \\
$ {\vec W}^{(54)} \, \leftrightarrow\, $ &$ h_4 E_{-\beta_1-\beta_2-\beta_3-\beta_4}+h_8 E_{\beta_1+\beta_2+\beta_3+
\beta_4} $   \\
$ {\vec W}^{(55)} \, \leftrightarrow\, $ &$   h_5 E_{-\beta_1-\beta_2-\beta_3-\beta_4-\beta_5}-h_8 E_{\beta_1+\beta_2+
\beta_3+\beta_4+\beta_5} $  \\
$ {\vec W}^{(56)} \, \leftrightarrow\, $ &$  h_6 E_{-\beta_1-\beta_2-\beta_3-\beta_4-\beta_5-\beta_6}+h_8 E_{\beta_1+
\beta_2+\beta_3+\beta_4+\beta_5+\beta_6}  $    \\
$ {\vec W}^{(29)} \, \leftrightarrow\, $ &$ -h_7 E_{-\beta_1-\beta_2-\beta_3-\beta_4-\beta_5-\beta_6
-\beta_7}-h_8 E_{\beta_1+\beta_2+\beta_3+\beta_4+\beta_5+\beta_6+\beta_7}   $ \\
$ {\vec W}^{(43)} \, \leftrightarrow\, $ &$ -h_3 E_{-\beta_3}-h_2
E_{\beta_3}  $    \\
$ {\vec W}^{(42)} \, \leftrightarrow\, $ &$ -h_4 E_{-\beta_3-\beta_4}+h_2
E_{\beta_3+\beta_4}   $  \\
$ {\vec W}^{(39)} \, \leftrightarrow\, $ &$ -h_5
E_{-\beta_3-\beta_4-\beta_5}-h_2 E_{\beta_3+\beta_4+\beta_5}  $    \\
$ {\vec W}^{(31)} \, \leftrightarrow\, $ &$ -h_6
E_{-\beta_3-\beta_4-\beta_5-\beta_6}+h_2 E_{\beta_3+\beta_4+\beta_5+\beta_6}  $  \\
$ {\vec W}^{(46)} \, \leftrightarrow\, $ &$ h_7
E_{-\beta_3-\beta_4-\beta_5-\beta_6-\beta_7}-h_2 E_{\beta_3+\beta_4+\beta_5+\beta_6+\beta_7}   $    \\
$ {\vec W}^{(44)} \, \leftrightarrow\, $ &$h_4 E_{-\beta_4}+h_3
E_{\beta_4}  $ \\
$ {\vec W}^{(40)} \, \leftrightarrow\, $ &$ h_5 E_{-\beta_4-\beta_5}-h_3 E_{\beta_4+\beta_5}  $    \\
$ {\vec W}^{(32)} \, \leftrightarrow\, $ &$  h_6
E_{-\beta_4-\beta_5-\beta_6}+ h_3 E_{\beta_4+\beta_5+\beta_6} $\\
$ {\vec W}^{(47)} \, \leftrightarrow\, $ &$  -h_7
E_{-\beta_4-\beta_5-\beta_6-\beta_7}- h_3 E_{\beta_4+\beta_5+\beta_6+\beta_7}  $    \\
$ {\vec W}^{(41)} \, \leftrightarrow\, $ &$ -h_5 E_{-\beta_5}-h_4
E_{\beta_5}  $ \\
$ {\vec W}^{(33)} \, \leftrightarrow\, $ &$   -h_6 E_{-\beta_5-\beta_6}+h_4
E_{\beta_5+\beta_6} $    \\
$ {\vec W}^{(48)} \, \leftrightarrow\, $ &$  h_7
E_{-\beta_5-\beta_6-\beta_7}-h_4
E_{\beta_5+\beta_6|\beta_7} $  \\
$ {\vec W}^{(34)} \, \leftrightarrow\, $ &$ h_6 E_{-\beta_6}+h_5
E_{\beta_6}   $    \\
$ {\vec W}^{(49)} \, \leftrightarrow\, $ &$ -h_7 E_{-\beta_6-\beta_7}-h_5
E_{\beta_6+\beta_7}  $ \\
$ {\vec W}^{(50)} \, \leftrightarrow\, $ &$  h_7 E_{-\beta_7}-h_6
E_{\beta_7} $ \\
\hline
\end{tabular}
\end{center}
\subsection{Comparison with Hull's results and description of the
algebras}
We can now compare the results we have obtained with those obtained
twelve years ago by Hull \cite{hull} and verify that the set of gaugings he had
obtained by his own method did exhaust  the set of possible N=8 theories.
The names we have given to the algebras we  have  retrieved in
our exhaustive search are the same names of his own algebras,
since the two sets coincide.
\par
To illustrate this one--to-one correspondence we briefly recall Hull's method of construction and
we describe the theories he obtained in addition to de Wit--Nicolai theory \cite{dwni}: this latter
corresponds to gauging the maximal compact
subalgebra of $SL(8,\IR)$ which is $SO(8)$ and
whose generators have the form $E_{\beta}-E_{-\beta}$, $\beta$
ranging on all the positive roots of $SL(8,\IR)$. This was the first gauging to be
studied  at the beginning of the
eighties. Later Hull found a new class of non--compact gaugings of
the $N=8$ theory consistent with the field dependent T--identities.
These new gauge algebras
$K_{\xi}(p,q),\, p+q=8$ were obtained from $SO(8)$ through the
conjugation of the latter by means of a one--parameter  $SL(8,\IR)$
transformation $E(t)$ generated by a suitable Cartan generator
$H_{(p,q)}$. For all the possible choices of the integers
$(p,q)\,\,p+q=8$, $H_{(p,q)}$ is defined as the Cartan generator
associated with the simple weight $\lambda_p$ corresponding to
$\beta_p$ and therefore it commutes with the $SL(p,\IR)$ and $SL(q,\IR)$
subalgebras of $SL(8,\IR)$ described by the root systems
$\{\gamma_1\}=\{\beta_1,\dots,\beta_{p-1}\}$ and $\{\gamma_2\}=
\{\beta_{p+1},\dots,\beta_7\}$
respectively. Therefore the conjugation through $E(t)=Exp(-tH_{(p,q)})$
leaves the subgroups $SO(p)\otimes SO(q) \subset SO(8)$
  invariant while it acts non trivially on
the remaining $p \times q$ generators in the ${\bf (p,q)}$
of the form $E_{\beta}-E_{-\beta}$,
in which $\beta$ varies on the positive roots containing $\beta_p$.
As far as the vector fields are
concerned, the $28$ electric vectors decompose with respect to
$SO(p)\otimes SO(q)$ in the following way:
\begin{equation}
{\bf 28}\rightarrow Adj(SO(p))+ Adj(SO(q))+{\bf(p,q)}
\label{28decomp}
\end{equation}
Let $a,b,c$ ($a<0$) be the eigenvalues of $H_{(p,q)}$ on $Adj(SO(p))$,
$Adj(SO(q))$ and ${\bf(p,q)}$ respectively.
Applying the transformation $E(t)$ to the vector fields as well yields
the following grading of the electric vectors (up to a redefinition of
the coupling constant):
\begin{eqnarray}
A^{AB}_{\mu}&\equiv &\{A^{p}_{\mu},A^{q}_{\mu},A^{(p,q)}_{\mu}\}\rightarrow
\{A^{p}_{\mu},\xi A^{q}_{\mu},\sqrt{\xi}A^{(p,q)}_{\mu}\}\nonumber\\
\xi \, &=&\, e^{(a-b)t}\,\,\,\,\,\sqrt{\xi}=e^{(a-c)t}
\end{eqnarray}
This grading can be transferred from the vector fields $A^{AB}_{\mu}$
to the corresponding $SO(8)$ generators $T^{AB}$. The
combined action of the conjugation by means of $E(t)$ and of the
aforementioned grading on the $SO(8)$ algebra yields an algebra
$K_{\xi}(p,q)$, which, once $t$ is extended analytically to complex
values,
is different from  $SO(8)$. In particular $K_{\xi}(p,q)$ has the
following form:
\begin{equation}
K_\xi(p,q)\,:\,\,\cases{$SO(8)$ & $\xi=1$\cr
SO(p,q) & $\xi=-1$ \cr
CSO(p,q)\equiv SO(p)\oplus {\bf (p,q)}_{nil} & $\xi=0 $ }
\label{kappacsi}
\end{equation}
 where ${\bf (p,q)}_{nil}$ consists of the shift operators $E_\beta $
 such that $\beta$ contains the simple root $\beta_p$.\par
 \par
 Our classification  includes all these algebras. In addition
 it also includes another kind of
 gauging corresponding to a different real form of $CSO(p,q)$,
namely of the form $SO(p_1,p_2)\oplus {\bf (p,q)}_{nil}\,\,p_1+p_2=p,\,
\,p+q=8$.
Indeed, by varying the possible combinations of the $h_i$ parameters we
 obtain algebras of the same dimensions as the $K_\xi(p,q)$, but
 corresponding to more general real forms. These additional real forms
 were also mentioned as a possibility by Hull in \cite{hull}, although
 he did not made an explicit construction of such cases.
 \par
 As many time stressed it appears from this classification that the
 $SO(8)$ Lie algebra, its non--compact real forms and the full
 set of all possible contractions thereof exhaust the set of $N=8$
 theories.
 \section{Comments on abelian gaugings and conclusions}
 We can now discuss the question of abelian gaugings that was the
 original motivation for the investigation we have performed.
 \par
 As it appears by inspection of the table in eq.\eqn{risulato}, among
 all the possible gaugings there is only one that is fully abelian,
 namely the case $CSO(1,7)$ corresponding to the choice $h_8=1,h_i=0
 (i \ne 8)$. The abelian character of this algebra is manifest from
 its signature that displays $28$ zeros and its the only one to do so.
 In this case it also appears that the number of generators that are
 actually gauged is $7$. This seems to be a confirmation of the
 prediction made in \cite{noi2}. There we had considered the
 possibility of gauging a subideal of the maximal abelian ideal of
 the solvable Lie algebra generating the $E_{7(7)}/SU(8)$ coset. We
 had come to the conclusion that the maximal gaugeable subideal was
 of dimension $7$. Such a conclusion was  simply based on the request
 that the ${\bf 28}$ representation should contain a number of singlets
 equal to the number of abelian generators to be gauged. However the
 algebraic $t$--identities had  not yet been taken into account so the
 existence of the gauging could not yet be established.
 Finding a unique $7$--dimensional abelian gauging is very
 encouraging but unfortunately it can be seen that this abelian
 algebra is not part of the maximal abelian ideal $\ID_6^+$ since in addition
 to positive roots $\{\beta_1+\beta_2,\dots,\beta_1+\beta_2+\beta_3+\beta_4+\beta_5+\beta_6+\beta_7\}\subset \ID_6^+$ it contains the root 
$\beta_1\in \ID_5^+$. Hence, although
 {\bf we have} an {\it abelian gauging}, it doesn't correspond to 
 gauging only {\it translational global symmetries} of the theory, the latter being associated with elements of the {\it maximal abelian ideal} ${\cal A}_7$. This makes the existence of flat directions
 doubtful and typically destroys our hopes of obtaining a spontaneous breaking of
 supersymmetry in Minkowski space. Yet in order to put such a conclusion on firm
 grounds one should make a systematic investigation of the
 scalar potentials generated by these gaugings and study their critical points.
 Such an analysis is at the moment non existing and would deserve further
 study.
 \par
 \subsection{Gauged supergravity and $p$--branes}
 More generally while the relation between the $M2$--brane solution and
 $SO(8)$--gauged supergravity has been established leading to a new
 exciting duality between Kaluza--Klein supergravity and world--volume
 conformal field theories \cite{maldapasto,goverh,holow,nairdaemi,lowersusy,zaffa}, such a relation for
 the non compact gaugings is so far neither established nor explored.
 It is clearly a stimulating question, but in order to address it one
 had to have an exhaustive classification of the available cases and
 this is what we have done in the present paper.

\vskip 10pt

{\bf \large Aknowledgments}
We are grateful to A. D'Adda and R. D'Auria for interesting and very useful 
discussions.



\begin{table}[ht]
\caption{{\bf
The  embedding matrix for positive roots of $SL(8,\IR)$:}}
\label{matap1}
\begin{center}
\begin{tabular}{|c|ccccccc|}
\hline
\hline
 $ \null $ & $ {E^{{{\rho }_1}}} $ & $ {E^{{{\rho }_2}}} $ & $
  {E^{{{\rho }_3}}} $ & $ {E^{{{\rho }_4}}} $ & $ {E^{{{\rho }_5}}} $ & $
  {E^{{{\rho }_6}}} $ & $ {E^{{{\rho }_7}}} $ \cr \hline $ {W_{29}} $ & $ 0 $ & $ 0 $ & $ 0 $ & $ 0
   $ & $ 0 $ & $ 0 $ & $ 0 $ \cr \hline $ {W_{30}} $ & $ 0 $ & $ 0 $ & $ 0 $ & $ 0 $ & $ -{h_1} $ & $ 0 $ & $ 0 $
   \cr \hline $
  {W_{31}} $ & $ {\ell_{17}} $ & $ 0 $ & $ 0 $ & $ 0 $ & $ -{\ell_3} $ & $ 0 $ & $ 0 $ \cr
  \hline $ {W_{32}} $ & $ 0
   $ & $ -{\ell_{17}} $ & $ 0 $ & $ 0 $ & $ -{\ell_5} $ & $ {\ell_{18}} $ & $ 0 $ \cr
   \hline $ {W_{33}} $ & $ 0 $ & $ 0
   $ & $ {\ell_{17}} $ & $ 0 $ & $ -{\ell_8} $ & $ 0 $ & $ -{\ell_{18}} $ \cr
   \hline $ {W_{34}} $ & $ 0 $ & $ 0 $ & $ 0
   $ & $ -{\ell_{17}} $ & $ -{\ell_{12}} $ & $ 0 $ & $ 0 $ \cr
   \hline $ {W_{35}} $ & $ -{h_1} $ & $ 0 $ & $ 0 $ & $ 0
   $ & $ 0 $ & $ 0 $ & $ 0 $ \cr
   \hline $ {W_{36}} $ & $ 0 $ & $ {h_1} $ & $ 0 $ & $ 0 $ & $ 0 $ & $ -{\ell_3} $ & $ 0
   $ \cr
   \hline $ {W_{37}} $ & $ 0 $ & $ 0 $ & $ -{h_1} $ & $ 0 $ & $ 0 $ & $ 0 $ & $ {\ell_3} $ \cr
   \hline $ {W_{38}} $ & $ 0
   $ & $ 0 $ & $ 0 $ & $ {h_1} $ & $ 0 $ & $ 0 $ & $ 0 $ \cr
   \hline $ {W_{39}} $ & $ {\ell_{12}} $ & $ 0 $ & $ 0 $ & $
  {\ell_3} $ & $ 0 $ & $ 0 $ & $ 0 $ \cr
  \hline $ {W_{40}} $ & $ 0 $ & $ -{\ell_{12}} $ & $ 0 $ & $ {\ell_5} $ & $ 0 $ & $
  {\ell_{13}} $ & $ 0 $ \cr
  \hline $ {W_{41}} $ & $ 0 $ & $ 0 $ & $ {\ell_{12}} $ & $ {\ell_8} $ & $ 0 $ & $ 0 $ & $
  -{\ell_{13}} $ \cr
  \hline $ {W_{42}} $ & $ {\ell_8} $ & $ 0 $ & $ -{\ell_3} $ & $ 0 $ & $ 0 $ & $ 0 $ & $ {h_2} $ \cr \hline $
  {W_{43}} $ & $ {\ell_5} $ & $ {\ell_3} $ & $ 0 $ & $ 0 $ & $ 0 $ & $ -{h_2} $ & $ 0 $ \cr
  \hline $ {W_{44}} $ & $ 0
   $ & $ -{\ell_8} $ & $ -{\ell_5} $ & $ 0 $ & $ 0 $ & $ {\ell_9} $ & $ {\ell_6} $ \cr \hline $ {W_{45}} $ & $ 0 $ & $ 0 $ & $ 0
   $ & $ 0 $ & $ 0 $ & $ 0 $ & $ 0 $ \cr
   \hline $ {W_{46}} $ & $ -{\ell_1} $ & $ 0 $ & $ 0 $ & $ 0 $ & $ 0 $ & $ 0 $ & $ 0
   $ \cr \hline $ {W_{47}} $ & $ 0 $ & $ {\ell_1} $ & $ 0 $ & $ 0 $ & $ 0 $ & $ -{\ell_2} $ & $ 0 $ \cr \hline $ {W_{48}} $ & $ 0
   $ & $ 0 $ & $ -{\ell_1} $ & $ 0 $ & $ 0 $ & $ 0 $ & $ {\ell_2} $ \cr \hline $ {W_{49}} $ & $ 0 $ & $ 0 $ & $ 0 $ & $ {\ell_1}
   $ & $ 0 $ & $ 0 $ & $ 0 $ \cr
   \hline $ {W_{50}} $ & $ 0 $ & $ 0 $ & $ 0 $ & $ 0 $ & $ -{\ell_1} $ & $ 0 $ & $ 0 $ \cr \hline $
  {W_{51}} $ & $ 0 $ & $ 0 $ & $ 0 $ & $ 0 $ & $ 0 $ & $ 0 $ & $ 0 $ \cr \hline $ {W_{52}} $ & $ {\ell_{23}} $ & $ 0
   $ & $ 0 $ & $ 0 $ & $ 0 $ & $ 0 $ & $ 0 $ \cr \hline $ {W_{53}} $ & $ 0 $ & $ -{\ell_{23}} $ & $ 0 $ & $ 0 $ & $ 0 $ & $
  -{\ell_{24}} $ & $ 0 $ \cr \hline $ {W_{54}} $ & $ 0 $ & $ 0 $ & $ {\ell_{23}} $ & $ 0 $ & $ 0 $ & $ 0 $ & $ {\ell_{24}}
   $ \cr \hline $ {W_{55}} $ & $ 0 $ & $ 0 $ & $ 0 $ & $ -{\ell_{23}} $ & $ 0 $ & $ 0 $ & $ 0 $ \cr
   \hline $ {W_{56}} $ & $ 0
   $ & $ 0 $ & $ 0 $ & $ 0 $ & $ {\ell_{23}} $ & $ 0 $ & $ 0 $ \cr
\hline
\end{tabular}
\end{center}
\end{table}

\begin{table}[ht]
\caption{{\bf
The  embedding matrix for positive roots of $SL(8,\IR)$
 (continued ) :}}
 \label{matap2}
\begin{center}
\begin{tabular}{|c|ccccccc|}
\hline
\hline
$ \null $ & $ {E^{{{\rho }_8}}} $ & $ {E^{{{\rho }_9}}} $ & $
  {E^{{{\rho }_{10}}}} $ & $ {E^{{{\rho }_{11}}}} $ & $ {E^{{{\rho }_{12}}}}
   $ & $ {E^{{{\rho }_{13}}}} $ & $ {E^{{{\rho }_{14}}}} $ \cr \hline $ {W_{29}} $ & $ 0 $ & $ 0
   $ & $ 0 $ & $ 0 $ & $ 0 $ & $ 0 $ & $ 0 $ \cr \hline $ {W_{30}} $ & $ 0 $ & $ {\ell_3} $ & $ 0 $ & $ 0 $ & $ -{\ell_5} $ & $
  0 $ & $ {\ell_8} $ \cr \hline $ {W_{31}} $ & $ 0 $ & $ {h_2} $ & $ 0 $ & $ 0 $ & $ -{\ell_6} $ & $ 0 $ & $ {\ell_9} $ \cr \hline $
  {W_{32}} $ & $ 0 $ & $ {\ell_6} $ & $ 0 $ & $ 0 $ & $ {h_3} $ & $ 0 $ & $ -{\ell_{10}} $ \cr \hline $ {W_{33}} $ & $
  0 $ & $ {\ell_9} $ & $ -{\ell_{19}} $ & $ 0 $ & $ {\ell_{10}} $ & $ 0 $ & $ {h_4} $ \cr \hline $ {W_{34}} $ & $
  {\ell_{18}} $ & $ {\ell_{13}} $ & $ 0 $ & $ {\ell_{19}} $ & $ {\ell_{14}} $ & $ {\ell_{20}} $ & $ {\ell_{15}}
   $ \cr \hline $ {W_{35}} $ & $ 0 $ & $ 0 $ & $ 0 $ & $ 0 $ & $ 0 $ & $ 0 $ & $ 0 $ \cr \hline $ {W_{36}} $ & $ 0 $ & $ 0 $ & $ 0
   $ & $ 0 $ & $ 0 $ & $ 0 $ & $ 0 $ \cr \hline $ {W_{37}} $ & $ 0 $ & $ 0 $ & $ -{\ell_5} $ & $ 0 $ & $ 0 $ & $ 0 $ & $ 0
   $ \cr \hline $ {W_{38}} $ & $ -{\ell_3} $ & $ 0 $ & $ 0 $ & $ {\ell_5} $ & $ 0 $ & $ -{\ell_8} $ & $ 0 $ \cr \hline $ {W_{39}}
   $ & $ -{h_2} $ & $ 0 $ & $ 0 $ & $ {\ell_6} $ & $ 0 $ & $ -{\ell_9} $ & $ 0 $ \cr \hline $ {W_{40}} $ & $ -{\ell_6} $ & $
  0 $ & $ 0 $ & $ -{h_3} $ & $ 0 $ & $ {\ell_{10}} $ & $ 0 $ \cr \hline $ {W_{41}} $ & $ -{\ell_9} $ & $ 0 $ & $
  -{\ell_{14}} $ & $ -{\ell_{10}} $ & $ 0 $ & $ -{h_4} $ & $ 0 $ \cr \hline $ {W_{42}} $ & $ 0 $ & $ 0 $ & $ -{\ell_6}
   $ & $ 0 $ & $ 0 $ & $ 0 $ & $ 0 $ \cr \hline $ {W_{43}} $ & $ 0 $ & $ 0 $ & $ 0 $ & $ 0 $ & $ 0 $ & $ 0 $ & $ 0 $ \cr \hline $
  {W_{44}} $ & $ 0 $ & $ 0 $ & $ {h_3} $ & $ 0 $ & $ 0 $ & $ 0 $ & $ 0 $ \cr \hline $ {W_{45}} $ & $ 0 $ & $ 0 $ & $ 0
   $ & $ 0 $ & $ 0 $ & $ 0 $ & $ 0 $ \cr \hline $ {W_{46}} $ & $ 0 $ & $ 0 $ & $ 0 $ & $ 0 $ & $ 0 $ & $ 0 $ & $ 0 $ \cr \hline $
  {W_{47}} $ & $ 0 $ & $ 0 $ & $ 0 $ & $ 0 $ & $ 0 $ & $ 0 $ & $ 0 $ \cr \hline $ {W_{48}} $ & $ 0 $ & $ 0 $ & $ -{\ell_4}
   $ & $ 0 $ & $ 0 $ & $ 0 $ & $ 0 $ \cr \hline $ {W_{49}} $ & $ -{\ell_2} $ & $ 0 $ & $ 0 $ & $ {\ell_4} $ & $ 0 $ & $
  -{\ell_7} $ & $ 0 $ \cr \hline $ {W_{50}} $ & $ 0 $ & $ {\ell_2} $ & $ 0 $ & $ 0 $ & $ -{\ell_4} $ & $ 0 $ & $ {\ell_7}
   $ \cr \hline $ {W_{51}} $ & $ 0 $ & $ 0 $ & $ 0 $ & $ 0 $ & $ 0 $ & $ 0 $ & $ 0 $ \cr \hline $ {W_{52}} $ & $ 0 $ & $ 0 $ & $ 0
   $ & $ 0 $ & $ 0 $ & $ 0 $ & $ 0 $ \cr \hline $ {W_{53}} $ & $ 0 $ & $ 0 $ & $ 0 $ & $ 0 $ & $ 0 $ & $ 0 $ & $ 0 $ \cr \hline $
  {W_{54}} $ & $ 0 $ & $ 0 $ & $ {\ell_{25}} $ & $ 0 $ & $ 0 $ & $ 0 $ & $ 0 $ \cr \hline $ {W_{55}} $ & $ -{\ell_{24}}
   $ & $ 0 $ & $ 0 $ & $ -{\ell_{25}} $ & $ 0 $ & $ -{\ell_{26}} $ & $ 0 $ \cr \hline $ {W_{56}} $ & $ 0 $ & $ {\ell_{24}}
   $ & $ 0 $ & $ 0 $ & $ {\ell_{25}} $ & $ 0 $ & $ {\ell_{26}} $ \cr
\hline
\end{tabular}
\end{center}
\end{table}

\begin{table}[ht]
\caption{{\bf
The  embedding matrix for positive roots of $SL(8,\IR)$
 (continued ) :}}
 \label{matap3}
\begin{center}
\begin{tabular}{|c|ccccccc|}
\hline
\hline
$ \null $ & $ {E^{{{\rho }_{15}}}} $ & $ {E^{{{\rho }_{16}}}} $ & $
  {E^{{{\rho }_{17}}}} $ & $ {E^{{{\rho }_{18}}}} $ & $ {E^{{{\rho }_{19}}}}
   $ & $ {E^{{{\rho }_{20}}}} $ & $ {E^{{{\rho }_{21}}}} $ \cr \hline $ {W_{29}} $ & $ 0 $ & $
  -{h_8} $ & $ -{\ell_{23}} $ & $ -{\ell_{24}} $ & $ -{\ell_{25}} $ & $ -{\ell_{26}} $ & $ -{\ell_{27}} $ \cr \hline $
  {W_{30}} $ & $ -{\ell_{12}} $ & $ 0 $ & $ 0 $ & $ 0 $ & $ 0 $ & $ 0 $ & $ 0 $ \cr \hline $ {W_{31}} $ & $
  -{\ell_{13}} $ & $ 0 $ & $ 0 $ & $ 0 $ & $ 0 $ & $ 0 $ & $ 0 $ \cr \hline $ {W_{32}} $ & $ {\ell_{14}} $ & $ 0 $ & $ 0
   $ & $ 0 $ & $ 0 $ & $ 0 $ & $ 0 $ \cr \hline $ {W_{33}} $ & $ -{\ell_{15}} $ & $ 0 $ & $ 0 $ & $ 0 $ & $ 0 $ & $ 0 $ & $ 0
   $ \cr \hline $ {W_{34}} $ & $ {h_5} $ & $ 0 $ & $ 0 $ & $ 0 $ & $ 0 $ & $ 0 $ & $ 0 $ \cr \hline $ {W_{35}} $ & $ 0 $ & $ 0
   $ & $ 0 $ & $ 0 $ & $ 0 $ & $ 0 $ & $ 0 $ \cr \hline $ {W_{36}} $ & $ 0 $ & $ 0 $ & $ 0 $ & $ 0 $ & $ 0 $ & $ 0 $ & $ 0
   $ \cr \hline $ {W_{37}} $ & $ 0 $ & $ 0 $ & $ 0 $ & $ 0 $ & $ 0 $ & $ 0 $ & $ 0 $ \cr \hline $ {W_{38}} $ & $ 0 $ & $ 0 $ & $ 0
   $ & $ 0 $ & $ 0 $ & $ 0 $ & $ 0 $ \cr \hline $ {W_{39}} $ & $ 0 $ & $ 0 $ & $ 0 $ & $ 0 $ & $ 0 $ & $ 0 $ & $ 0 $ \cr \hline $
  {W_{40}} $ & $ 0 $ & $ 0 $ & $ 0 $ & $ 0 $ & $ 0 $ & $ 0 $ & $ 0 $ \cr \hline $ {W_{41}} $ & $ 0 $ & $ 0 $ & $ 0 $ & $ 0
   $ & $ 0 $ & $ 0 $ & $ 0 $ \cr \hline $ {W_{42}} $ & $ 0 $ & $ 0 $ & $ 0 $ & $ 0 $ & $ 0 $ & $ 0 $ & $ 0 $ \cr \hline $ {W_{43}}
   $ & $ 0 $ & $ 0 $ & $ 0 $ & $ 0 $ & $ 0 $ & $ 0 $ & $ 0 $ \cr \hline $ {W_{44}} $ & $ 0 $ & $ 0 $ & $ 0 $ & $ 0 $ & $ 0
   $ & $ 0 $ & $ 0 $ \cr \hline $ {W_{45}} $ & $ 0 $ & $ -{\ell_{23}} $ & $ {h_1} $ & $ -{\ell_3} $ & $ {\ell_5} $ & $
  -{\ell_8} $ & $ {\ell_{12}} $ \cr \hline $ {W_{46}} $ & $ 0 $ & $ {\ell_{24}} $ & $ {\ell_3} $ & $ -{h_2} $ & $ {\ell_6}
   $ & $ -{\ell_9} $ & $ {\ell_{13}} $ \cr \hline $ {W_{47}} $ & $ 0 $ & $ -{\ell_{25}} $ & $ {\ell_5} $ & $ -{\ell_6} $ & $
  -{h_3} $ & $ {\ell_{10}} $ & $ -{\ell_{14}} $ \cr \hline $ {W_{48}} $ & $ 0 $ & $ {\ell_{26}} $ & $ {\ell_8} $ & $
  -{\ell_9} $ & $ -{\ell_{10}} $ & $ -{h_4} $ & $ {\ell_{15}} $ \cr \hline $ {W_{49}} $ & $ 0 $ & $ -{\ell_{27}} $ & $
  {\ell_{12}} $ & $ -{\ell_{13}} $ & $ -{\ell_{14}} $ & $ -{\ell_{15}} $ & $ -{h_5} $ \cr \hline $ {W_{50}} $ & $
  -{\ell_{11}} $ & $ -{\ell_{28}} $ & $ {\ell_{17}} $ & $ -{\ell_{18}} $ & $ -{\ell_{19}} $ & $ -{\ell_{20}}
   $ & $ -{\ell_{21}} $ \cr \hline $ {W_{51}} $ & $ 0 $ & $ 0 $ & $ 0 $ & $ 0 $ & $ 0 $ & $ 0 $ & $ 0 $ \cr \hline $ {W_{52}}
   $ & $ 0 $ & $ 0 $ & $ 0 $ & $ 0 $ & $ 0 $ & $ 0 $ & $ 0 $ \cr \hline $ {W_{53}} $ & $ 0 $ & $ 0 $ & $ 0 $ & $ 0 $ & $ 0
   $ & $ 0 $ & $ 0 $ \cr \hline $ {W_{54}} $ & $ 0 $ & $ 0 $ & $ 0 $ & $ 0 $ & $ 0 $ & $ 0 $ & $ 0 $ \cr \hline $ {W_{55}} $ & $ 0
   $ & $ 0 $ & $ 0 $ & $ 0 $ & $ 0 $ & $ 0 $ & $ 0 $ \cr \hline $ {W_{56}} $ & $ {\ell_{27}} $ & $ 0 $ & $ 0 $ & $ 0 $ & $ 0
   $ & $ 0 $ & $ 0 $ \cr
\hline
\end{tabular}
\end{center}
\end{table}

\begin{table}[ht]
\caption{{\bf
The  embedding matrix for positive roots of $SL(8,\IR)$
 (continued ) :}}
 \label{matap4}
\begin{center}
\begin{tabular}{|c|ccccccc|}
\hline
\hline
$ \null $ & $ {E^{{{\rho }_{22}}}} $ & $ {E^{{{\rho }_{23}}}} $ & $
  {E^{{{\rho }_{24}}}} $ & $ {E^{{{\rho }_{25}}}} $ & $ {E^{{{\rho }_{26}}}}
   $ & $ {E^{{{\rho }_{27}}}} $ & $ {E^{{{\rho }_{28}}}} $ \cr \hline $ {W_{29}} $ & $ 0 $ & $ 0
   $ & $ 0 $ & $ 0 $ & $ 0 $ & $ 0 $ & $ {\ell_{28}} $ \cr \hline $ {W_{30}} $ & $ -{\ell_{28}} $ & $ 0 $ & $ 0 $ & $ 0
   $ & $ 0 $ & $ {\ell_{23}} $ & $ 0 $ \cr \hline $ {W_{31}} $ & $ 0 $ & $ -{\ell_{28}} $ & $ 0 $ & $ 0 $ & $ 0 $ & $
  -{\ell_{24}} $ & $ 0 $ \cr \hline $ {W_{32}} $ & $ 0 $ & $ 0 $ & $ {\ell_{28}} $ & $ 0 $ & $ 0 $ & $ {\ell_{25}} $ & $ 0
   $ \cr \hline $ {W_{33}} $ & $ 0 $ & $ 0 $ & $ 0 $ & $ -{\ell_{28}} $ & $ 0 $ & $ -{\ell_{26}} $ & $ 0 $ \cr \hline $
  {W_{34}} $ & $ 0 $ & $ 0 $ & $ 0 $ & $ 0 $ & $ {\ell_{28}} $ & $ {\ell_{27}} $ & $ 0 $ \cr \hline $ {W_{35}} $ & $
  {\ell_{24}} $ & $ {\ell_{23}} $ & $ 0 $ & $ 0 $ & $ 0 $ & $ 0 $ & $ 0 $ \cr \hline $ {W_{36}} $ & $ -{\ell_{25}} $ & $ 0
   $ & $ -{\ell_{23}} $ & $ 0 $ & $ 0 $ & $ 0 $ & $ 0 $ \cr \hline $ {W_{37}} $ & $ {\ell_{26}} $ & $ 0 $ & $ 0 $ & $
  {\ell_{23}} $ & $ 0 $ & $ 0 $ & $ 0 $ \cr \hline $ {W_{38}} $ & $ -{\ell_{27}} $ & $ 0 $ & $ 0 $ & $ 0 $ & $
  -{\ell_{23}} $ & $ 0 $ & $ 0 $ \cr \hline $ {W_{39}} $ & $ 0 $ & $ -{\ell_{27}} $ & $ 0 $ & $ 0 $ & $ {\ell_{24}} $ & $
  0 $ & $ 0 $ \cr \hline $ {W_{40}} $ & $ 0 $ & $ 0 $ & $ {\ell_{27}} $ & $ 0 $ & $ -{\ell_{25}} $ & $ 0 $ & $ 0 $ \cr \hline $
  {W_{41}} $ & $ 0 $ & $ 0 $ & $ 0 $ & $ -{\ell_{27}} $ & $ {\ell_{26}} $ & $ 0 $ & $ 0 $ \cr \hline $ {W_{42}} $ & $ 0
   $ & $ {\ell_{26}} $ & $ 0 $ & $ -{\ell_{24}} $ & $ 0 $ & $ 0 $ & $ 0 $ \cr \hline $ {W_{43}} $ & $ 0 $ & $ -{\ell_{25}}
   $ & $ {\ell_{24}} $ & $ 0 $ & $ 0 $ & $ 0 $ & $ 0 $ \cr \hline $ {W_{44}} $ & $ 0 $ & $ 0 $ & $ -{\ell_{26}} $ & $
  {\ell_{25}} $ & $ 0 $ & $ 0 $ & $ 0 $ \cr \hline $ {W_{45}} $ & $ -{\ell_{22}} $ & $ 0 $ & $ 0 $ & $ 0 $ & $ 0 $ & $ 0
   $ & $ -{\ell_{17}} $ \cr \hline $ {W_{46}} $ & $ 0 $ & $ -{\ell_{22}} $ & $ 0 $ & $ 0 $ & $ 0 $ & $ 0 $ & $
  -{\ell_{18}} $ \cr \hline $ {W_{47}} $ & $ 0 $ & $ 0 $ & $ {\ell_{22}} $ & $ 0 $ & $ 0 $ & $ 0 $ & $ {\ell_{19}} $ \cr \hline $
  {W_{48}} $ & $ 0 $ & $ 0 $ & $ 0 $ & $ -{\ell_{22}} $ & $ 0 $ & $ 0 $ & $ -{\ell_{20}} $ \cr \hline $ {W_{49}} $ & $
  0 $ & $ 0 $ & $ 0 $ & $ 0 $ & $ {\ell_{22}} $ & $ 0 $ & $ {\ell_{21}} $ \cr \hline $ {W_{50}} $ & $ 0 $ & $ 0 $ & $ 0
   $ & $ 0 $ & $ 0 $ & $ -{\ell_{22}} $ & $ -{h_6} $ \cr \hline $ {W_{51}} $ & $ {h_8} $ & $ 0 $ & $ 0 $ & $ 0 $ & $ 0
   $ & $ 0 $ & $ 0 $ \cr \hline $ {W_{52}} $ & $ 0 $ & $ {h_8} $ & $ 0 $ & $ 0 $ & $ 0 $ & $ 0 $ & $ 0 $ \cr \hline $ {W_{53}}
   $ & $ 0 $ & $ 0 $ & $ -{h_8} $ & $ 0 $ & $ 0 $ & $ 0 $ & $ 0 $ \cr \hline $ {W_{54}} $ & $ 0 $ & $ 0 $ & $ 0 $ & $
  {h_8} $ & $ 0 $ & $ 0 $ & $ 0 $ \cr \hline $ {W_{55}} $ & $ 0 $ & $ 0 $ & $ 0 $ & $ 0 $ & $ -{h_8} $ & $ 0 $ & $ 0
   $ \cr \hline $ {W_{56}} $ & $ 0 $ & $ 0 $ & $ 0 $ & $ 0 $ & $ 0 $ & $ {h_8} $ & $ 0 $ \cr
\hline
\end{tabular}
\end{center}
\end{table}

\begin{table}[ht]
\caption{{\bf
The  embedding matrix for negative roots of $SL(8,\IR)$
 :}}
 \label{matam1}
\begin{center}
\begin{tabular}{|c|ccccccc|}
\hline
\hline
 $ \null $ & $ {E^{-{{\rho }_1}}} $ & $ {E^{-{{\rho }_2}}} $ & $
  {E^{-{{\rho }_3}}} $ & $ {E^{-{{\rho }_4}}} $ & $ {E^{-{{\rho }_5}}} $ & $
  {E^{-{{\rho }_6}}} $ & $ {E^{-{{\rho }_7}}} $ \cr \hline $ {W_{29}} $ & $ 0 $ & $ 0 $ & $ 0 $ & $
  0 $ & $ 0 $ & $ 0 $ & $ 0 $ \cr \hline $ {W_{30}} $ & $ {\ell_{18}} $ & $ {\ell_{19}} $ & $ {\ell_{20}} $ & $
  {\ell_{21}} $ & $ {h_6} $ & $ 0 $ & $ 0 $ \cr \hline $ {W_{31}} $ & $ 0 $ & $ 0 $ & $ 0 $ & $ 0 $ & $ 0 $ & $
  -{\ell_{19}} $ & $ -{\ell_{20}} $ \cr \hline $ {W_{32}} $ & $ 0 $ & $ 0 $ & $ 0 $ & $ 0 $ & $ 0 $ & $ 0 $ & $ 0 $ \cr \hline $
  {W_{33}} $ & $ 0 $ & $ 0 $ & $ 0 $ & $ 0 $ & $ 0 $ & $ 0 $ & $ 0 $ \cr \hline $ {W_{34}} $ & $ 0 $ & $ 0 $ & $ 0 $ & $ 0
   $ & $ 0 $ & $ 0 $ & $ 0 $ \cr \hline $ {W_{35}} $ & $ {h_2} $ & $ -{\ell_6} $ & $ {\ell_9} $ & $ -{\ell_{13}} $ & $
  {\ell_{18}} $ & $ -{\ell_5} $ & $ {\ell_8} $ \cr \hline $ {W_{36}} $ & $ {\ell_6} $ & $ {h_3} $ & $ -{\ell_{10}} $ & $
  {\ell_{14}} $ & $ -{\ell_{19}} $ & $ 0 $ & $ 0 $ \cr \hline $ {W_{37}} $ & $ {\ell_9} $ & $ {\ell_{10}} $ & $ {h_4}
   $ & $ -{\ell_{15}} $ & $ {\ell_{20}} $ & $ 0 $ & $ 0 $ \cr \hline $ {W_{38}} $ & $ {\ell_{13}} $ & $ {\ell_{14}} $ & $
  {\ell_{15}} $ & $ {h_5} $ & $ -{\ell_{21}} $ & $ 0 $ & $ 0 $ \cr \hline $ {W_{39}} $ & $ 0 $ & $ 0 $ & $ 0 $ & $ 0
   $ & $ 0 $ & $ -{\ell_{14}} $ & $ -{\ell_{15}} $ \cr \hline $ {W_{40}} $ & $ 0 $ & $ 0 $ & $ 0 $ & $ 0 $ & $ 0 $ & $ 0
   $ & $ 0 $ \cr \hline $ {W_{41}} $ & $ 0 $ & $ 0 $ & $ 0 $ & $ 0 $ & $ 0 $ & $ 0 $ & $ 0 $ \cr \hline $ {W_{42}} $ & $ 0 $ & $ 0
   $ & $ 0 $ & $ 0 $ & $ 0 $ & $ -{\ell_{10}} $ & $ -{h_4} $ \cr \hline $ {W_{43}} $ & $ 0 $ & $ 0 $ & $ 0 $ & $ 0 $ & $ 0
   $ & $ -{h_3} $ & $ {\ell_{10}} $ \cr \hline $ {W_{44}} $ & $ 0 $ & $ 0 $ & $ 0 $ & $ 0 $ & $ 0 $ & $ 0 $ & $ 0 $ \cr \hline $
  {W_{45}} $ & $ -{\ell_2} $ & $ {\ell_4} $ & $ -{\ell_7} $ & $ {\ell_{11}} $ & $ -{\ell_{16}} $ & $ 0 $ & $ 0
   $ \cr \hline $ {W_{46}} $ & $ 0 $ & $ 0 $ & $ 0 $ & $ 0 $ & $ 0 $ & $ -{\ell_4} $ & $ {\ell_7} $ \cr \hline $ {W_{47}} $ & $ 0
   $ & $ 0 $ & $ 0 $ & $ 0 $ & $ 0 $ & $ 0 $ & $ 0 $ \cr \hline $ {W_{48}} $ & $ 0 $ & $ 0 $ & $ 0 $ & $ 0 $ & $ 0 $ & $ 0
   $ & $ 0 $ \cr \hline $ {W_{49}} $ & $ 0 $ & $ 0 $ & $ 0 $ & $ 0 $ & $ 0 $ & $ 0 $ & $ 0 $ \cr \hline $ {W_{50}} $ & $ 0 $ & $ 0
   $ & $ 0 $ & $ 0 $ & $ 0 $ & $ 0 $ & $ 0 $ \cr \hline $ {W_{51}} $ & $ -{\ell_{24}} $ & $ -{\ell_{25}} $ & $
  -{\ell_{26}} $ & $ -{\ell_{27}} $ & $ {\ell_{28}} $ & $ 0 $ & $ 0 $ \cr \hline $ {W_{52}} $ & $ 0 $ & $ 0 $ & $ 0 $ & $
  0 $ & $ 0 $ & $ {\ell_{25}} $ & $ {\ell_{26}} $ \cr \hline $ {W_{53}} $ & $ 0 $ & $ 0 $ & $ 0 $ & $ 0 $ & $ 0 $ & $ 0
   $ & $ 0 $ \cr \hline $ {W_{54}} $ & $ 0 $ & $ 0 $ & $ 0 $ & $ 0 $ & $ 0 $ & $ 0 $ & $ 0 $ \cr \hline $ {W_{55}} $ & $ 0 $ & $ 0
   $ & $ 0 $ & $ 0 $ & $ 0 $ & $ 0 $ & $ 0 $ \cr \hline $ {W_{56}} $ & $ 0 $ & $ 0 $ & $ 0 $ & $ 0 $ & $ 0 $ & $ 0 $ & $ 0
   $ \cr
\hline
\end{tabular}
\end{center}
\end{table}

\begin{table}[ht]
\caption{{\bf
The  embedding matrix for negative roots of $SL(8,\IR)$
 (continued) :}}
 \label{matam2}
\begin{center}
\begin{tabular}{|c|ccccccc|}
\hline
\hline
$ \null $ & $ {E^{-{{\rho }_8}}} $ & $ {E^{-{{\rho }_9}}} $ & $
  {E^{-{{\rho }_{10}}}} $ & $ {E^{-{{\rho }_{11}}}} $ & $
  {E^{-{{\rho }_{12}}}} $ & $ {E^{-{{\rho }_{13}}}} $ & $
  {E^{-{{\rho }_{14}}}} $ \cr \hline $ {W_{29}} $ & $ 0 $ & $ 0 $ & $ 0 $ & $ 0 $ & $ 0 $ & $ 0 $ & $ 0 $ \cr \hline $
  {W_{30}} $ & $ 0 $ & $ 0 $ & $ 0 $ & $ 0 $ & $ 0 $ & $ 0 $ & $ 0 $ \cr \hline $ {W_{31}} $ & $ -{\ell_{21}} $ & $
  -{h_6} $ & $ 0 $ & $ 0 $ & $ 0 $ & $ 0 $ & $ 0 $ \cr \hline $ {W_{32}} $ & $ 0 $ & $ 0 $ & $ {\ell_{20}} $ & $
  {\ell_{21}} $ & $ {h_6} $ & $ 0 $ & $ 0 $ \cr \hline $ {W_{33}} $ & $ 0 $ & $ 0 $ & $ 0 $ & $ 0 $ & $ 0 $ & $
  -{\ell_{21}} $ & $ -{h_6} $ \cr \hline $ {W_{34}} $ & $ 0 $ & $ 0 $ & $ 0 $ & $ 0 $ & $ 0 $ & $ 0 $ & $ 0 $ \cr \hline $
  {W_{35}} $ & $ -{\ell_{12}} $ & $ {\ell_{17}} $ & $ 0 $ & $ 0 $ & $ 0 $ & $ 0 $ & $ 0 $ \cr \hline $ {W_{36}} $ & $ 0
   $ & $ 0 $ & $ -{\ell_8} $ & $ {\ell_{12}} $ & $ -{\ell_{17}} $ & $ 0 $ & $ 0 $ \cr \hline $ {W_{37}} $ & $ 0 $ & $ 0
   $ & $ 0 $ & $ 0 $ & $ 0 $ & $ -{\ell_{12}} $ & $ {\ell_{17}} $ \cr \hline $ {W_{38}} $ & $ 0 $ & $ 0 $ & $ 0 $ & $ 0
   $ & $ 0 $ & $ 0 $ & $ 0 $ \cr \hline $ {W_{39}} $ & $ -{h_5} $ & $ {\ell_{21}} $ & $ 0 $ & $ 0 $ & $ 0 $ & $ 0 $ & $ 0
   $ \cr \hline $ {W_{40}} $ & $ 0 $ & $ 0 $ & $ {\ell_{15}} $ & $ {h_5} $ & $ -{\ell_{21}} $ & $ 0 $ & $ 0 $ \cr \hline $
  {W_{41}} $ & $ 0 $ & $ 0 $ & $ 0 $ & $ 0 $ & $ 0 $ & $ -{h_5} $ & $ {\ell_{21}} $ \cr \hline $ {W_{42}} $ & $
  {\ell_{15}} $ & $ -{\ell_{20}} $ & $ 0 $ & $ 0 $ & $ 0 $ & $ -{\ell_{13}} $ & $ {\ell_{18}} $ \cr \hline $ {W_{43}}
   $ & $ -{\ell_{14}} $ & $ {\ell_{19}} $ & $ -{\ell_9} $ & $ {\ell_{13}} $ & $ -{\ell_{18}} $ & $ 0 $ & $ 0 $ \cr \hline $
  {W_{44}} $ & $ 0 $ & $ 0 $ & $ {h_4} $ & $ -{\ell_{15}} $ & $ {\ell_{20}} $ & $ {\ell_{14}} $ & $
  -{\ell_{19}} $ \cr \hline $ {W_{45}} $ & $ 0 $ & $ 0 $ & $ 0 $ & $ 0 $ & $ 0 $ & $ 0 $ & $ 0 $ \cr \hline $ {W_{46}} $ & $
  -{\ell_{11}} $ & $ {\ell_{16}} $ & $ 0 $ & $ 0 $ & $ 0 $ & $ 0 $ & $ 0 $ \cr \hline $ {W_{47}} $ & $ 0 $ & $ 0 $ & $
  -{\ell_7} $ & $ {\ell_{11}} $ & $ -{\ell_{16}} $ & $ 0 $ & $ 0 $ \cr \hline $ {W_{48}} $ & $ 0 $ & $ 0 $ & $ 0 $ & $ 0
   $ & $ 0 $ & $ -{\ell_{11}} $ & $ {\ell_{16}} $ \cr \hline $ {W_{49}} $ & $ 0 $ & $ 0 $ & $ 0 $ & $ 0 $ & $ 0 $ & $ 0
   $ & $ 0 $ \cr \hline $ {W_{50}} $ & $ 0 $ & $ 0 $ & $ 0 $ & $ 0 $ & $ 0 $ & $ 0 $ & $ 0 $ \cr \hline $ {W_{51}} $ & $ 0 $ & $ 0
   $ & $ 0 $ & $ 0 $ & $ 0 $ & $ 0 $ & $ 0 $ \cr \hline $ {W_{52}} $ & $ {\ell_{27}} $ & $ -{\ell_{28}} $ & $ 0 $ & $ 0
   $ & $ 0 $ & $ 0 $ & $ 0 $ \cr \hline $ {W_{53}} $ & $ 0 $ & $ 0 $ & $ -{\ell_{26}} $ & $ -{\ell_{27}} $ & $ {\ell_{28}}
   $ & $ 0 $ & $ 0 $ \cr \hline $ {W_{54}} $ & $ 0 $ & $ 0 $ & $ 0 $ & $ 0 $ & $ 0 $ & $ {\ell_{27}} $ & $ -{\ell_{28}}
   $ \cr \hline $ {W_{55}} $ & $ 0 $ & $ 0 $ & $ 0 $ & $ 0 $ & $ 0 $ & $ 0 $ & $ 0 $ \cr \hline $ {W_{56}} $ & $ 0 $ & $ 0 $ & $ 0
   $ & $ 0 $ & $ 0 $ & $ 0 $ & $ 0 $ \cr
   \hline
\end{tabular}
\end{center}
\end{table}

\begin{table}[ht]
\caption{{\bf
The  embedding matrix for negative roots of $SL(8,\IR)$
 (continued) :}}
 \label{matam3}
\begin{center}
\begin{tabular}{|c|ccccccc|}
\hline
\hline
$ \null $ & $ {E^{-{{\rho }_{15}}}} $ & $ {E^{-{{\rho }_{16}}}} $ & $
  {E^{-{{\rho }_{17}}}} $ & $ {E^{-{{\rho }_{18}}}} $ & $
  {E^{-{{\rho }_{19}}}} $ & $ {E^{-{{\rho }_{20}}}} $ & $
  {E^{-{{\rho }_{21}}}} $ \cr \hline $ {W_{29}} $ & $ 0 $ & $ -{h_7} $ & $ 0 $ & $ 0 $ & $ 0 $ & $ 0 $ & $ 0
   $ \cr \hline $ {W_{30}} $ & $ 0 $ & $ 0 $ & $ {\ell_{16}} $ & $ 0 $ & $ 0 $ & $ 0 $ & $ 0 $ \cr \hline $ {W_{31}} $ & $ 0
   $ & $ 0 $ & $ 0 $ & $ -{\ell_{16}} $ & $ 0 $ & $ 0 $ & $ 0 $ \cr \hline $ {W_{32}} $ & $ 0 $ & $ 0 $ & $ 0 $ & $ 0 $ & $
  {\ell_{16}} $ & $ 0 $ & $ 0 $ \cr \hline $ {W_{33}} $ & $ 0 $ & $ 0 $ & $ 0 $ & $ 0 $ & $ 0 $ & $ -{\ell_{16}} $ & $ 0
   $ \cr \hline $ {W_{34}} $ & $ {h_6} $ & $ 0 $ & $ 0 $ & $ 0 $ & $ 0 $ & $ 0 $ & $ {\ell_{16}} $ \cr \hline $ {W_{35}} $ & $
  0 $ & $ 0 $ & $ {\ell_2} $ & $ {\ell_1} $ & $ 0 $ & $ 0 $ & $ 0 $ \cr \hline $ {W_{36}} $ & $ 0 $ & $ 0 $ & $ {\ell_4} $ & $ 0
   $ & $ -{\ell_1} $ & $ 0 $ & $ 0 $ \cr \hline $ {W_{37}} $ & $ 0 $ & $ 0 $ & $ {\ell_7} $ & $ 0 $ & $ 0 $ & $ {\ell_1} $ & $ 0
   $ \cr \hline $ {W_{38}} $ & $ -{\ell_{17}} $ & $ 0 $ & $ {\ell_{11}} $ & $ 0 $ & $ 0 $ & $ 0 $ & $ -{\ell_1} $ \cr \hline $
  {W_{39}} $ & $ -{\ell_{18}} $ & $ 0 $ & $ 0 $ & $ -{\ell_{11}} $ & $ 0 $ & $ 0 $ & $ -{\ell_2} $ \cr \hline $
  {W_{40}} $ & $ {\ell_{19}} $ & $ 0 $ & $ 0 $ & $ 0 $ & $ {\ell_{11}} $ & $ 0 $ & $ -{\ell_4} $ \cr \hline $ {W_{41}}
   $ & $ -{\ell_{20}} $ & $ 0 $ & $ 0 $ & $ 0 $ & $ 0 $ & $ -{\ell_{11}} $ & $ -{\ell_7} $ \cr \hline $ {W_{42}} $ & $ 0
   $ & $ 0 $ & $ 0 $ & $ -{\ell_7} $ & $ 0 $ & $ {\ell_2} $ & $ 0 $ \cr \hline $ {W_{43}} $ & $ 0 $ & $ 0 $ & $ 0 $ & $
  -{\ell_4} $ & $ -{\ell_2} $ & $ 0 $ & $ 0 $ \cr \hline $ {W_{44}} $ & $ 0 $ & $ 0 $ & $ 0 $ & $ 0 $ & $ {\ell_7} $ & $
  {\ell_4} $ & $ 0 $ \cr \hline $ {W_{45}} $ & $ 0 $ & $ 0 $ & $ -{h_7} $ & $ 0 $ & $ 0 $ & $ 0 $ & $ 0 $ \cr \hline $ {W_{46}}
   $ & $ 0 $ & $ 0 $ & $ 0 $ & $ {h_7} $ & $ 0 $ & $ 0 $ & $ 0 $ \cr \hline $ {W_{47}} $ & $ 0 $ & $ 0 $ & $ 0 $ & $ 0 $ & $
  -{h_7} $ & $ 0 $ & $ 0 $ \cr \hline $ {W_{48}} $ & $ 0 $ & $ 0 $ & $ 0 $ & $ 0 $ & $ 0 $ & $ {h_7} $ & $ 0 $ \cr \hline $
  {W_{49}} $ & $ -{\ell_{16}} $ & $ 0 $ & $ 0 $ & $ 0 $ & $ 0 $ & $ 0 $ & $ -{h_7} $ \cr \hline $ {W_{50}} $ & $ 0
   $ & $ 0 $ & $ 0 $ & $ 0 $ & $ 0 $ & $ 0 $ & $ 0 $ \cr \hline $ {W_{51}} $ & $ 0 $ & $ {\ell_1} $ & $ -{\ell_{22}} $ & $ 0
   $ & $ 0 $ & $ 0 $ & $ 0 $ \cr \hline $ {W_{52}} $ & $ 0 $ & $ {\ell_2} $ & $ 0 $ & $ {\ell_{22}} $ & $ 0 $ & $ 0 $ & $ 0
   $ \cr \hline $ {W_{53}} $ & $ 0 $ & $ {\ell_4} $ & $ 0 $ & $ 0 $ & $ -{\ell_{22}} $ & $ 0 $ & $ 0 $ \cr \hline $ {W_{54}}
   $ & $ 0 $ & $ {\ell_7} $ & $ 0 $ & $ 0 $ & $ 0 $ & $ {\ell_{22}} $ & $ 0 $ \cr \hline $ {W_{55}} $ & $ {\ell_{28}} $ & $
  {\ell_{11}} $ & $ 0 $ & $ 0 $ & $ 0 $ & $ 0 $ & $ -{\ell_{22}} $ \cr \hline $ {W_{56}} $ & $ 0 $ & $ {\ell_{16}} $ & $ 0
   $ & $ 0 $ & $ 0 $ & $ 0 $ & $ 0 $ \cr
\hline
\end{tabular}
\end{center}
\end{table}

\begin{table}[ht]
\caption{{\bf
The  embedding matrix for negative roots of $SL(8,\IR)$
 (continued) :}}
 \label{matam4}
\begin{center}
\begin{tabular}{|c|ccccccc|}
\hline
\hline
$ \null $ & $ {{{E^{-{{\rho }_{22}}}}}}  $ & $
  {{{E^{-{{\rho }_{23}}}}}} $ & $
  {{{E^{-{{\rho }_{24}}}}}} $ & $
  {{{E^{-{{\rho }_{25}}}}}} $ & $
  {{{E^{-{{\rho }_{26}}}}}} $ & $
  {{{E^{-{{\rho }_{27}}}}}} $ & $
  {{{E^{-{{\rho }_{28}}}}}} $ \cr \hline $ {W_{29}} $ & $ -{\ell_1} $ & $
  -{\ell_2} $ & $ {\ell_4} $ & $ -{\ell_7} $ & $ {\ell_{11}} $ & $ -{\ell_{16}} $ & $ 0 $ \cr \hline $ {W_{30}} $ & $ 0
   $ & $ 0 $ & $ 0 $ & $ 0 $ & $ 0 $ & $ 0 $ & $ {\ell_1} $ \cr \hline $ {W_{31}} $ & $ 0 $ & $ 0 $ & $ 0 $ & $ 0 $ & $ 0 $ & $
  0 $ & $ {\ell_2} $ \cr \hline $ {W_{32}} $ & $ 0 $ & $ 0 $ & $ 0 $ & $ 0 $ & $ 0 $ & $ 0 $ & $ {\ell_4} $ \cr \hline $ {W_{33}}
   $ & $ 0 $ & $ 0 $ & $ 0 $ & $ 0 $ & $ 0 $ & $ 0 $ & $ {\ell_7} $ \cr \hline $ {W_{34}} $ & $ 0 $ & $ 0 $ & $ 0 $ & $ 0 $ & $
  0 $ & $ 0 $ & $ {\ell_{11}} $ \cr \hline $ {W_{35}} $ & $ 0 $ & $ 0 $ & $ 0 $ & $ 0 $ & $ 0 $ & $ 0 $ & $ 0 $ \cr \hline $
  {W_{36}} $ & $ 0 $ & $ 0 $ & $ 0 $ & $ 0 $ & $ 0 $ & $ 0 $ & $ 0 $ \cr \hline $ {W_{37}} $ & $ 0 $ & $ 0 $ & $ 0 $ & $ 0
   $ & $ 0 $ & $ 0 $ & $ 0 $ \cr \hline $ {W_{38}} $ & $ 0 $ & $ 0 $ & $ 0 $ & $ 0 $ & $ 0 $ & $ 0 $ & $ 0 $ \cr \hline $ {W_{39}}
   $ & $ 0 $ & $ 0 $ & $ 0 $ & $ 0 $ & $ 0 $ & $ 0 $ & $ 0 $ \cr \hline $ {W_{40}} $ & $ 0 $ & $ 0 $ & $ 0 $ & $ 0 $ & $ 0
   $ & $ 0 $ & $ 0 $ \cr \hline $ {W_{41}} $ & $ 0 $ & $ 0 $ & $ 0 $ & $ 0 $ & $ 0 $ & $ 0 $ & $ 0 $ \cr \hline $ {W_{42}} $ & $ 0
   $ & $ 0 $ & $ 0 $ & $ 0 $ & $ 0 $ & $ 0 $ & $ 0 $ \cr \hline $ {W_{43}} $ & $ 0 $ & $ 0 $ & $ 0 $ & $ 0 $ & $ 0 $ & $ 0
   $ & $ 0 $ \cr \hline $ {W_{44}} $ & $ 0 $ & $ 0 $ & $ 0 $ & $ 0 $ & $ 0 $ & $ 0 $ & $ 0 $ \cr \hline $ {W_{45}} $ & $ 0 $ & $ 0
   $ & $ 0 $ & $ 0 $ & $ 0 $ & $ 0 $ & $ 0 $ \cr \hline $ {W_{46}} $ & $ 0 $ & $ 0 $ & $ 0 $ & $ 0 $ & $ 0 $ & $ 0 $ & $ 0
   $ \cr \hline $ {W_{47}} $ & $ 0 $ & $ 0 $ & $ 0 $ & $ 0 $ & $ 0 $ & $ 0 $ & $ 0 $ \cr \hline $ {W_{48}} $ & $ 0 $ & $ 0 $ & $ 0
   $ & $ 0 $ & $ 0 $ & $ 0 $ & $ 0 $ \cr \hline $ {W_{49}} $ & $ 0 $ & $ 0 $ & $ 0 $ & $ 0 $ & $ 0 $ & $ 0 $ & $ 0 $ \cr \hline $
  {W_{50}} $ & $ 0 $ & $ 0 $ & $ 0 $ & $ 0 $ & $ 0 $ & $ 0 $ & $ {h_7} $ \cr \hline $ {W_{51}} $ & $ {h_1} $ & $
  {\ell_3} $ & $ -{\ell_5} $ & $ {\ell_8} $ & $ -{\ell_{12}} $ & $ {\ell_{17}} $ & $ 0 $ \cr \hline $ {W_{52}} $ & $ {\ell_3}
   $ & $ {h_2} $ & $ -{\ell_6} $ & $ {\ell_9} $ & $ -{\ell_{13}} $ & $ {\ell_{18}} $ & $ 0 $ \cr \hline $ {W_{53}} $ & $
  {\ell_5} $ & $ {\ell_6} $ & $ {h_3} $ & $ -{\ell_{10}} $ & $ {\ell_{14}} $ & $ -{\ell_{19}} $ & $ 0 $ \cr \hline $
  {W_{54}} $ & $ {\ell_8} $ & $ {\ell_9} $ & $ {\ell_{10}} $ & $ {h_4} $ & $ -{\ell_{15}} $ & $ {\ell_{20}} $ & $
  0 $ \cr \hline $ {W_{55}} $ & $ {\ell_{12}} $ & $ {\ell_{13}} $ & $ {\ell_{14}} $ & $ {\ell_{15}} $ & $ {h_5} $ & $
  -{\ell_{21}} $ & $ 0 $ \cr \hline $ {W_{56}} $ & $ {\ell_{17}} $ & $ {\ell_{18}} $ & $ {\ell_{19}} $ & $ {\ell_{20}}
   $ & $ {\ell_{21}} $ & $ {h_6} $ & $ {\ell_{22}} $ \cr
   \hline
\end{tabular}
\end{center}
\end{table}

\begin{table}[ht]
\caption{{\bf
The  embedding matrix for Cartan generators of $SL(8,\IR)$
   :}}
 \label{matacar}
\begin{center}
\begin{tabular}{|c|ccccccc|}
\hline
\hline
$ \null $ & $ {C_1} $ & $ {C_2} $ & $ {C_3} $ & $ {C_4} $ & $ {C_5} $ & $ {C_6} $ & $ {C_7} $ \cr \hline $
  {W_{29}} $ & $ {\ell_{22}} $ & $ 2\,{\ell_{22}} $ & $ 3\,{\ell_{22}} $ & $ 4\,{\ell_{22}} $ & $
  2\,{\ell_{22}} $ & $ 3\,{\ell_{22}} $ & $ 2\,{\ell_{22}} $ \cr \hline $ {W_{30}} $ & $ -{\ell_{17}} $ & $
  -{\ell_{17}} $ & $ -{\ell_{17}} $ & $ -{\ell_{17}} $ & $ 0 $ & $ -{\ell_{17}} $ & $ 0 $ \cr \hline $ {W_{31}} $ & $ 0
   $ & $ -{\ell_{18}} $ & $ -{\ell_{18}} $ & $ -{\ell_{18}} $ & $ 0 $ & $ -{\ell_{18}} $ & $ 0 $ \cr \hline $ {W_{32}}
   $ & $ 0 $ & $ 0 $ & $ {\ell_{19}} $ & $ {\ell_{19}} $ & $ 0 $ & $ {\ell_{19}} $ & $ 0 $ \cr \hline $ {W_{33}} $ & $ 0
   $ & $ 0 $ & $ 0 $ & $ -{\ell_{20}} $ & $ 0 $ & $ -{\ell_{20}} $ & $ 0 $ \cr \hline $ {W_{34}} $ & $ 0 $ & $ 0 $ & $ 0
   $ & $ 0 $ & $ 0 $ & $ {\ell_{21}} $ & $ 0 $ \cr \hline $ {W_{35}} $ & $ -{\ell_3} $ & $ 0 $ & $ 0 $ & $ 0 $ & $ 0 $ & $ 0
   $ & $ 0 $ \cr \hline $ {W_{36}} $ & $ -{\ell_5} $ & $ -{\ell_5} $ & $ 0 $ & $ 0 $ & $ 0 $ & $ 0 $ & $ 0 $ \cr \hline $ {W_{37}}
   $ & $ -{\ell_8} $ & $ -{\ell_8} $ & $ -{\ell_8} $ & $ 0 $ & $ 0 $ & $ 0 $ & $ 0 $ \cr \hline $ {W_{38}} $ & $ -{\ell_{12}}
   $ & $ -{\ell_{12}} $ & $ -{\ell_{12}} $ & $ -{\ell_{12}} $ & $ 0 $ & $ 0 $ & $ 0 $ \cr \hline $ {W_{39}} $ & $ 0 $ & $
  -{\ell_{13}} $ & $ -{\ell_{13}} $ & $ -{\ell_{13}} $ & $ 0 $ & $ 0 $ & $ 0 $ \cr \hline $ {W_{40}} $ & $ 0 $ & $ 0
   $ & $ {\ell_{14}} $ & $ {\ell_{14}} $ & $ 0 $ & $ 0 $ & $ 0 $ \cr \hline $ {W_{41}} $ & $ 0 $ & $ 0 $ & $ 0 $ & $
  -{\ell_{15}} $ & $ 0 $ & $ 0 $ & $ 0 $ \cr \hline $ {W_{42}} $ & $ 0 $ & $ -{\ell_9} $ & $ -{\ell_9} $ & $ 0 $ & $ 0 $ & $
  0 $ & $ 0 $ \cr \hline $ {W_{43}} $ & $ 0 $ & $ -{\ell_6} $ & $ 0 $ & $ 0 $ & $ 0 $ & $ 0 $ & $ 0 $ \cr \hline $ {W_{44}} $ & $
  0 $ & $ 0 $ & $ {\ell_{10}} $ & $ 0 $ & $ 0 $ & $ 0 $ & $ 0 $ \cr \hline $ {W_{45}} $ & $ {\ell_1} $ & $ {\ell_1} $ & $
  {\ell_1} $ & $ {\ell_1} $ & $ 0 $ & $ {\ell_1} $ & $ {\ell_1} $ \cr \hline $ {W_{46}} $ & $ 0 $ & $ {\ell_2} $ & $ {\ell_2}
   $ & $ {\ell_2} $ & $ 0 $ & $ {\ell_2} $ & $ {\ell_2} $ \cr \hline $ {W_{47}} $ & $ 0 $ & $ 0 $ & $ {\ell_4} $ & $ {\ell_4}
   $ & $ 0 $ & $ {\ell_4} $ & $ {\ell_4} $ \cr \hline $ {W_{48}} $ & $ 0 $ & $ 0 $ & $ 0 $ & $ {\ell_7} $ & $ 0 $ & $ {\ell_7}
   $ & $ {\ell_7} $ \cr \hline $ {W_{49}} $ & $ 0 $ & $ 0 $ & $ 0 $ & $ 0 $ & $ 0 $ & $ {\ell_{11}} $ & $ {\ell_{11}} $ \cr \hline $
  {W_{50}} $ & $ 0 $ & $ 0 $ & $ 0 $ & $ 0 $ & $ 0 $ & $ 0 $ & $ {\ell_{16}} $ \cr \hline $ {W_{51}} $ & $ 0 $ & $
  {\ell_{23}} $ & $ 2\,{\ell_{23}} $ & $ 3\,{\ell_{23}} $ & $ 2\,{\ell_{23}} $ & $ 2\,{\ell_{23}} $ & $
  {\ell_{23}} $ \cr \hline $ {W_{52}} $ & $ -{\ell_{24}} $ & $ -{\ell_{24}} $ & $ -2\,{\ell_{24}} $ & $
  -3\,{\ell_{24}} $ & $ -2\,{\ell_{24}} $ & $ -2\,{\ell_{24}} $ & $ -{\ell_{24}} $ \cr \hline $ {W_{53}} $ & $
  {\ell_{25}} $ & $ 2\,{\ell_{25}} $ & $ 2\,{\ell_{25}} $ & $ 3\,{\ell_{25}} $ & $ 2\,{\ell_{25}} $ & $
  2\,{\ell_{25}} $ & $ {\ell_{25}} $ \cr \hline $ {W_{54}} $ & $ -{\ell_{26}} $ & $ -2\,{\ell_{26}} $ & $
  -3\,{\ell_{26}} $ & $ -3\,{\ell_{26}} $ & $ -2\,{\ell_{26}} $ & $ -2\,{\ell_{26}} $ & $ -{\ell_{26}}
   $ \cr \hline $ {W_{55}} $ & $ {\ell_{27}} $ & $ 2\,{\ell_{27}} $ & $ 3\,{\ell_{27}} $ & $ 4\,{\ell_{27}} $ & $
  2\,{\ell_{27}} $ & $ 2\,{\ell_{27}} $ & $ {\ell_{27}} $ \cr \hline $ {W_{56}} $ & $ {\ell_{28}} $ & $
  2\,{\ell_{28}} $ & $ 3\,{\ell_{28}} $ & $ 4\,{\ell_{28}} $ & $ 2\,{\ell_{28}} $ & $ 3\,{\ell_{28}} $ & $
  {\ell_{28}} $ \cr
\hline
\end{tabular}
\end{center}
\end{table}

\end{document}